\documentclass[twocolumn]{aastex631}

\usepackage{graphics}
\usepackage{graphicx}
\usepackage{amsmath}
\usepackage{xcolor}

\DeclareMathOperator{\STM}{\mathbf{\Phi}}
\DeclareMathOperator{\state}{\bar{X}}
\DeclareMathOperator{\dstate}{\dot{\bar{X}}}
\DeclareMathOperator{\Jacobian}{\mathbf{A}}
\DeclareMathOperator{\Fconst}{\mathbf{G}(\state_d)}
\DeclareMathOperator{\Xd}{\state_d}

\DeclareMathOperator{\acr}{a\textsubscript{c}}
\DeclareMathOperator{\acLW}{a\textsubscript{c,LW}}
\DeclareMathOperator{\acHW}{a\textsubscript{c,HW}}
\DeclareMathOperator{\acDB}{a\textsubscript{c,DB}}
\DeclareMathOperator{\acQ}{a\textsubscript{c,Q}}

\begin{document}

\title{A Dynamical Systems Approach to the \\ Theory of Circumbinary Orbits in the Circular Restricted Problem}

\shorttitle{Theory of Circumbinary Orbits}
\shortauthors{Langford and Weiss}

\submitjournal{The Astronomical Journal} 
\accepted{January 30, 2023}
\received{November 28, 2022}

\author[0000-0001-5312-649X]{Andrew Langford}
\altaffiliation{NSF Graduate Research Fellow}
\affiliation{Department of Physics and Astronomy, University of Notre Dame, Notre Dame, IN, 46556, USA}
\affiliation{Institute of Astronomy, University of Cambridge, Madingley Road, Cambridge CB3 0HA, UK}

\author[0000-0002-3725-3058]{Lauren M. Weiss}
\affiliation{Department of Physics and Astronomy, University of Notre Dame, Notre Dame, IN, 46556, USA}

\correspondingauthor{Andrew Langford}
\email{aml99@cam.ac.uk}


\begin{abstract}

To better understand the orbital dynamics of exoplanets around close binary stars, i.e., circumbinary planets (CBPs), we applied techniques from dynamical systems theory to a physically motivated set of solutions in the Circular Restricted Three-Body Problem (CR3BP). We applied Floquet theory to characterize the linear dynamical behavior -- static, oscillatory, or exponential -- surrounding planar circumbinary periodic trajectories (limit cycles). We computed prograde and retrograde limit cycles and analyzed their geometries, stability bifurcations, and dynamical structures. Orbit and stability calculations are exact computations in the CR3BP and reproducible through the open-source Python package \href{https://github.com/alangfor/pyraa}{\texttt{pyraa}}. The \href{https://doi.org/10.5281/zenodo.7532982}{periodic trajectories} produce a set of non-crossing, dynamically cool circumbinary orbits conducive to planetesimal growth. For mass ratios $\mu \in [0.01, 0.50]$ we found recurring features in the prograde families. These features include: (1) an innermost near-circular trajectory, inside which solutions have resonant geometries, (2) an innermost stable trajectory ($\acr \approx 1.61 - 1.85 \, a_\textrm{bin}$) characterized by a tangent bifurcating limit cycle, and (3) a region of dynamical instability ($a \approx 2.1 \ a_\textrm{bin}; \Delta a \approx 0.1 \ a_\textrm{bin}$), the exclusion zone, bounded by a pair of critically stable trajectories -- bifurcating limit cycles. The exterior boundary of the exclusion zone is consistent with prior determinations of $\acr$ around a circular binary. We validate our analytic results with N-body simulations  
 and apply them to the Pluto-Charon system. The absence of detected CBPs in the inner stable region, between the prograde exclusion zone and $\acr$, suggests that the exclusion zone may inhibit the inward migration of CBPs.

\end{abstract}

\section{Introduction}

\begin{deluxetable*}{lcccccccc}[t]
\tablecaption{Methods and concepts comparison between selected investigations of circumbinary orbital dynamics. The checkmark symbol denotes a feature present within the work. The times symbol denotes an absence of a feature. \label{Table: Methods}}
\tablehead{\colhead{Paper} & \colhead{Model} & \colhead{Periodic orbits} & \colhead{Floquet theory} & \colhead{Stability limit} & \colhead{Retrograde} & \colhead{Out-of-plane} & \colhead{Polar orbits}} 
\startdata
\cite{Stromgren1922} & CR3BP  & $\checkmark$ & $\times$   & $\times$       & $\checkmark$ & $\times$    & $\times$ \\
\cite{Holman1999}    & ER3BP  & $\times$   & $\times$     & $\checkmark$   & $\times$    & $\times$   & $\times$   \\
\cite{Doolin2011}    & ER3BP  & $\times$     & $\times$     & $\checkmark$   & $\checkmark$  & $\checkmark$ & $\checkmark$  \\
\cite{BosanacHowellFischbach2015}   & CR3BP  & $\checkmark$ & $\checkmark$ & $\times$   & $\times$    & $\times$   & $\times$   \\
\cite{Quarles2018}   & N-body & $\times$     & $\times$     & $\checkmark$   & $\times$    & $\checkmark$ & $\checkmark$  \\
\cite{Chen2020}      & N-body & $\times$     & $\times$   & $\checkmark$   & $\checkmark$  & $\checkmark$ & $\checkmark$  \\
This Paper           & CR3BP  & $\checkmark$ & $\checkmark$ & $\checkmark$ & $\checkmark$  & $\checkmark$  & $\times$   \\
\enddata
\tablecomments{The metrics selected are for purposes of relating to this investigation and do not represent the total methods and analysis of any given work.}
\end{deluxetable*}

The current census of over 5200 exoplanets reveals a variety of planetary system configurations, many of which are distinct from the Solar System \citep{ExoplanetArchive}.  Although close binary stars with separations $a\textsubscript{bin} < 1 $ AU comprise ten billion stellar systems in the Milky Way Galaxy \citep{Raghavan2010}, most confirmed exoplanets inhabit single-star systems or binaries with long orbital periods, $P\textsubscript{bin} > 100$ years. Because the detection of circumbinary planets (CBPs) is rare and observational follow-up of such planetary systems often poses challenges, this intricate -- yet possibly abundant -- architecture is less well understood than systems orbiting single stars. To date, 14 transiting CBPs have been confirmed from Kepler and TESS photometry, despite an unfavorable selection bias imposed by pipeline rejection for eclipsing binaries \citep{Kostov2021, Jenkins2010}. CBPs have also been detected through direct imaging techniques \citep{Janson2021}. With Kepler data, \citet{Armstrong2014} determined that the occurrence of CBPs with R\textsubscript{p} $>$ 6 R\textsubscript{$\Earth$} with orbital periods $<$ 300 days is $10.0 ^{+18}_{-6.5} \% \ (95\% \ \textrm{conf.})$---comparable to the occurrence of giant planets in single star systems \citep{Fulton2021}. Orbital trajectories surrounding the binary may result in ejection from the system as a result of the nonlinear and chaotic dynamics produced by the binary stars' gravitational potential. Intriguingly, many of these CBPs orbit close to their dynamical stability limit as determined by long ($10^4 \, P_\textrm{bin})$ numerical integration \citet{Holman1999}. \citet{Li2016} demonstrated that this CBP pile-up is not likely a result of observational bias.  

Despite these challenges in detecting CBPs, the hierarchical three-body interaction between the planet(s) and the two stars typically yields more precise ephemerides and dynamical masses than what can be determined for planets orbiting single stars \citep{WinnFabrycky2015}. The stellar binary also tends to stabilize a CBP's spin obliquity -- a feature conducive to long-term climate regulation \citep{Deitrick2018, ChenLiTao2022}. When observed, close binary systems represent some of the most information-rich laboratories to test the robustness of theories for planetary formation and evolution. 

The theoretical basis for how circumbinary gravitational interactions impact planetary evolution is an active area of study. CBPs exist within a dynamically sensitive gravitational environment as two stars orbit interior to the planet's trajectory. The Keplerian framework originally developed for Solar System dynamics does not apply as systematically to circumbinary as to single star orbits. The non-integrable nature and nonlinear sensitivity of the three-body problem pose challenges to producing generalized, precise predictions about the dynamics and stability of the system. 

Two methods that are readily applicable to the orbital dynamics around a binary star system are long-duration numerical integration and dynamical systems theory. \citet{Holman1999} provided one of the first long-duration numerical experiments to determine broad patterns of bounded circumbinary trajectories. Similar to more recent implementations, they tested the survivability of Keplerian circular initial conditions ($v(r) \propto \sqrt{1/r}$) to probe feasible locations for a planet's trajectory around a binary star \citep{Doolin2011, Quarles2018, Chen2020}. Advances in this approach have come from longer integration times, finer sampling, inclination consideration and planet/disk mass modeling. The bounded trajectories of long-term integrations are often presented by their osculating Keplerian Elements and initial distance from the binary. Through this method, three stable circumbinary orbital configurations: retrograde planar, prograde planar, and polar have been established \citep{Doolin2011, Chen2019}.

The dynamical systems approach aims to understand the natural dynamical structures of a nonlinear solution space by studying its fundamental solutions, i.e. periodic orbits, equilibria, invariant stable and unstable manifolds. In contrast to perturbative analytic approaches to circumbinary dynamics that predict motion in terms of two-body solutions and osculating Keplerian elements \citep{LeePeale2006, LeungLee2013, GeorgakarakosEggl2015, SutherlandKratter2019}, dynamical systems theory packs the three-body complexity directly into the system's fundamental solutions and then analyzes those solutions. The approach originates from mathematics developed by \citet{Poincare1892}'s study of the three-body problem. Early work in the field of dynamical systems applied to orbital mechanics recognized the \textit{practical} and \textit{mathematical} significance of periodic trajectories -- as both guaranteed bounded numerical results and probes to the underlying solution space \citep{Szeb1967}. Numerical integration by human computers at the Copenhagen Observatory solved the first continuous sets or \textit{families} of periodic trajectories in retrograde and prograde motion around an equal-mass binary, consequently termed the Copenhagen Problem \citep{Stromgren1922, Stromgren1938, Szeb1967}. In the late 20\textsuperscript{th} century, the astrodynamics community developed numerical techniques for computing families of periodic trajectories to understand the nonlinear solution space and stability of CR3BP trajectories \citep{Breakwell1979, Howell1984}. In this modern approach, the stability in a region of state space is calculated with respect to the linear dynamics surrounding nearby nonlinear periodic solution through Floquet theory \citep{Floquet1883}. \citet{BosanacHowellFischbach2015} introduced these modern techniques for computing families of prograde circumbinary periodic trajectories and assessing their stability through Floquet theory over a wide range of stellar mass ratios $(\mu \in [10^{-6}, 0.50])$. In addition to exact numerical results, the strength of the dynamical systems approach comes from understanding the underlying dynamical structures of an idealized nonlinear solution space that tend to persist within real physical systems \citep{Bosanac2016}.

The location of the innermost trajectory for a planet around a binary star is an ongoing question in dynamical astronomy. Interior to this trajectory, the gravitational environment produces chaotic dynamics prone to ejecting small masses from the system. This dynamical constraint likely shapes existing CBP architectures and is distinct from the ejection mechanisms in single star systems -- although destabilizing planet-planet resonances can also be induced in circumbinary systems \citep{SutherlandKratter2019}. Consequently, determining the closest stable, and thus observable, orbit has been a common endeavor in investigations that employ long-term numerical integrations \citep{Holman1999, Doolin2011, Quarles2018, Chen2020}. Intriguingly, the majority of CBP discoveries have orbits near the innermost stable distance as determined by these long-term numerical integrations \cite{Armstrong2014, WinnFabrycky2015, Li2016}.  

To develop a theory for the orbital dynamics of circumbinary exoplanets, we use techniques from modern dynamical systems theory (\S\ref{sec:dynamical_background}). Our approach builds on the computations of \citet{BosanacHowellFischbach2015}, leveraging families of periodic solutions (limit cycles) to understand how the solution space and stability evolve within systems and across mass ratios. We first demonstrate this approach in the special case of an equal-mass binary (\S\ref{Sec: CP}), and then generalized to a range of binary mass ratios (\S\ref{Sec: MassRatios}). To contextualize our calculations, we compare determinations of critically stable trajectories with past works (\S\ref{Sec: a_cr}). We discuss the implications our computations have with respect to theories for CBP formation and evolution (\S \ref{Sec: DST4PF}). Table \ref{Table: Methods} compares our analysis to other investigations of circumbinary orbital dynamics. 

This paper is intended to be the first in a series to develop the theory and application of a dynamical systems approach to describe multi-body gravitational environments in astrophysical systems. 

\section{Dynamical Background and Theory}
\label{sec:dynamical_background}

A major challenge of the three-body problem is that a set of generalized fundamental solutions is not available in closed form. By instead modeling circumbinary planetary motion in a Circular Restricted Three-body Problem (CR3BP), where the two stars have mass and the planet is treated as a test particle, we can numerically solve the system's fundamental, non-static, bounded solutions, continuous families of limit cycles, and assess their intrinsic stability properties using Floquet theory. 

\begin{figure}[b]
    \centering
    \includegraphics[width = \linewidth]{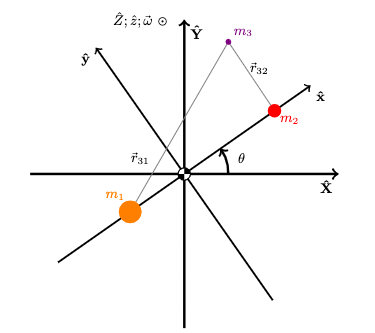}
    \caption{Schematic of the CR3BP. A test particle ($m_3=0$) follows a trajectory under the gravitational influence of two massive bodies ($m_1 >0 $, $m_2 > 0$), where the orbits of $m_1$ and $m_2$ are circular. The synodic reference frame ($\hat{x},\hat{y},\hat{z}$) rotates at rate $d\vec{\theta}/dt = \vec{\omega}$ with respect to the inertial frame ($\hat{X},\hat{Y},\hat{Z}$) in the common $\hat{z}/\hat{Z}$ direction, with the origin at the barycenter.}
    \label{fig:CR3BP_geom}
\end{figure}

\subsection{Model: Circular Restricted Three-Body Problem} \label{Sec: Model}

The CR3BP describes the motion of a test particle ($m_3 = 0$) under the gravitational influence of two massive bodies ($m_1$, $m_2 > 0$), where $m_1$ and $m_2$ orbit the system barycenter in circular motion \citep{Euler1772}. The geometry of the CR3BP is shown in Figure \ref{fig:CR3BP_geom}.  The equations of motion (EOM) can be reduced from 18 to 6 dimensions by studying the dynamics of $m_3$ in a synodic reference frame that co-rotates with the binary, effectively removing the motions of $m_1$ and $m_2$. The six-dimensional synodic EOM have five fixed points, famously known as the Lagrange Points, and one integral constant, the Jacobi Constant \citep{Lagrange1772, Jacobi1836}. Most notably for this investigation, the non-autonomous, i.e., time-independent, EOM permit continuous families of periodic solutions \citep{Poincare1892}. The remainder of \S\ref{sec:dynamical_background} reviews the pertinent CR3BP analytical theory used in this work. See \citet{Szeb1967, Murray2000, Short2010, Pavlak2010, Bosanac2016, Gupta2020, Boudad2022} for more careful treatment of the analytical theory. 

\subsubsection{Equations of Motion} \label{Sec: EOM}

The CR3BP has one parameter, the mass ratio,
\begin{equation} \label{Eq: mu_def}
    \mu := \frac{m_2}{m_1 + m_2} \in (0, 0.5],
\end{equation}
evaluated as the mass partitioning of the two primary masses. The mass ratio determines the dynamical solution space and generalizes solutions to physical systems. The CR3BP is often expressed in terms of non-dimensional equations of motion (EOM), which are related to dimensional EOM via characteristic physical quantities (Table \ref{tab:ndim_quants}). In this paper, we use non-dimensional, synodic frame coordinates unless otherwise stated. The positions of the primary masses are static in the synodic frame and located such that the center of mass remains fixed at the origin, $\vec{r}_1 = -\mu\hat{x}$ and $\vec{r}_2 = (1-\mu)\hat{x}$.

\begin{deluxetable}{ccl}[t]
\tablecaption{Table of dimensional and non-dimensional values with their characteristic quantities in the CR3BP. Dimensional values are obtained by multiplying non-dimensional values by the respective characteristic quantity. \label{tab:ndim_quants}}
\tablehead{\colhead{Dimensional} & \colhead{Non-Dimensional} & \colhead{Characteristic} } 
\startdata
$m_1$ & $1-\mu$ & $m^* = m_1 + m_2$ \\ 
$m_2$ & $\mu$ & $m^* = m_1 + m_2$ \\ 
$a_{bin}$ & $1$ & $l^* = |\vec{r}_1 - \vec{r}_2|$ \\ 
$P_{bin}$ & $2\pi$ & $t^*$ = $\sqrt{\frac{l^{*3}}{Gm^*}}$ \\
\enddata
\end{deluxetable}

By convention, a \textit{pseudo-potential} function is defined as, 
\begin{equation} \label{Eq: Psuedo_Pot}
    U^*(x,y,z) = \frac{n^2}{2}(x^2 + y^2) + \frac{1-\mu}{|\vec{r}_{31}|} + \frac{\mu}{|\vec{r}_{32}|} 
\end{equation}
where $n$ is the mean motion of the binary and is set to $n=1$ in the CR3BP.  The pseudo-potential incorporates the gravitational potentials of $m_1$, $m_2$, and the effective potential of centrifugal force. The CR3BP equations of motion are 

\begin{subequations} \label{EOM_Psued}
\renewcommand{\theequation}{\theparentequation.\arabic{equation}}
\begin{align}
    \ddot{x} &= (\nabla U^*)_x + 2n\dot{y} \\
    \ddot{y} &= (\nabla U^*)_y - 2n\dot{x} \\
    \ddot{z} &= (\nabla U^*)_z
\end{align}
\end{subequations}

The CR3BP's set of three, second-order, nonlinear ordinary differential equations can be uncoupled into a set of six first-order nonlinear ODE's that satisfy the relation 
\begin{equation} \label{Eq: F(X)}
    \mathbf{F}(\bar{X}) = \dot{\bar{X}} ,
\end{equation}
where $\state = [x, y, z, \dot{x}, \dot{y}, \dot{z}]^T$ is the instantaneous state of the particle and $\dstate$ is given by the CR3BP equations of motion (Equation \ref{EOM_Psued}). Given a set of six initial conditions, $\state(t_0)$, $\state$ may be evaluated at a given time $t$ by numerically integrating the vector function $\mathbf{F}(\bar{X})$ from $t_0$ to $t$. 

\subsubsection{Jacobi Constant} \label{Sec: CJ}

The CR3BP equations of motion in Equation \ref{EOM_Psued} possess time translational symmetry, i.e., no explicit time dependence. From Noether’s Theorem,
we should therefore expect an energy-like quantity to be conserved \citep{Noether1918}. This invariant quantity is by convention the Jacobi Constant,

\begin{equation} \label{Eq: CJ}
    C_J  := 2U^*(x,y,z) - (\dot{x}^2 + \dot{y}^2 + \dot{z}^2).
\end{equation}

Note the relation of the Jacobi Constant to the pseudo-Hamiltonian, $H^* = -\frac{1}{2}C_J$. Large positive values of the Jacobi Constant refer to ``low energy'' trajectories.

Since a particle's Jacobi Constant is invariant along a ballistic trajectory, asserting real-valued velocity components results in equations for the zero-velocity curves and surfaces – boundaries to regions in configuration space $(\hat{x}, \hat{y}, \hat{z})$ that are not-accessible to $m_3$. The Jacobi constant also reduces the system's degrees of freedom from six to five. In cases of planar dynamics, $z = \dot{z} = 0$, the problem is reduced to three degrees of freedom. 

\subsubsection{Jacobian and State Transition Matrix} \label{Sec: STM}

In a dynamical system, the Jacobian Matrix, 
\begin{equation}
    \Jacobian(\state) := \frac{\partial \mathbf{F}(\state)}{\partial \state}
\end{equation}

provides information about the linear dynamics surrounding a particular state. In the CR3BP, differentiating Equation \ref{Eq: F(X)} with respect to the state vector, $\state$, results in,

\begin{equation}
    \Jacobian_{CR3BP}(\state) = 
    \begin{bmatrix}
    
    0 & 0 & 0 & 1 & 0 & 0  \\
    0 & 0 & 0 & 0 & 1 & 0  \\
    0 & 0 & 0 & 0 & 0 & 1  \\
    U^*_{xx} & U^*_{xy} & U^*_{xz} & 0 & 2 & 0 \\
    U^*_{yx} & U^*_{yy} & U^*_{yz} & -2 & 2 & 0 \\
    U^*_{zx} & U^*_{zy} & U^*_{zz} & 0 & 0 & 0 \\
    
    \end{bmatrix}.
\end{equation}
where $U^*_{ij}$ refers to the second partial derivative of the pseudo-potential (Equation \ref{Eq: Psuedo_Pot}) with respect to $i$ and $j$.

Computing the eigenvalues and eigenvectors of $\mathbf{A}_{CR3BP}(\bar{X})$ at the fixed points, $\{\bar{X} \ | \ \mathbf{F}(\bar{X}) = 0\}$, allows for their surrounding linear dynamics to be classified. Similar analysis of non-static bounded solutions, i.e. limit cycles, begins with computing the State Transition Matrix (STM), $\mathbf{\Phi}$.  Consider a vector function, $\mathbf{F}(\state) = \dstate$, evaluated at $\state^* + \delta \state$ as a Taylor expansion around $\state^*$,

\begin{equation} \label{Eq: STM_dev0}
    \mathbf{F}(\state^* + \delta \state) = \dstate^* + \Jacobian(\state^*)\delta\state + O(\delta\state^2)
\end{equation}
where $\mathbf{A}$ is the Jacobian matrix. To first order, Equation \ref{Eq: STM_dev0} becomes
\begin{equation}
    \delta\dstate = \mathbf{A}(\state ^*)\delta\state
\end{equation}
which has a general solution

\begin{equation} \label{Eq: STM}
    \delta\state(t) = \STM(t, t_0)\delta\state(t_0)
\end{equation}
where $\STM$ describes the linear relation between two states at times $t_0$ at $t$.  A few pertinent properties of $\STM$ are 
\begin{subequations} \label{Eq: STM_Props}
\renewcommand{\theequation}{\theparentequation.\arabic{equation}}
\begin{align}
    \STM(t_0 + \delta t, t_0) & = \frac{\partial \state(t_0 + \delta t)}{\partial \state(t_0)} \label{STM1}\\ 
    \dot{\STM}(t_0, t_0) &= \Jacobian(t_0)\state(t_0) \label{STM2}\\
    \STM(t_N, t_0) &= \Pi^{N-1}_0 \STM(t_{i+1}, t_i) \label{STM3}.
\end{align}
\end{subequations}

Equations \ref{STM1}, \ref{STM2}, and \ref{STM3} describe a linear map that carries information about how initial perturbations in a given state component are propagated into the state vector at a given time. As a result, the STM carries intrinsic dynamical information about a reference trajectory through state space. It also serves as a valuable tool for solving periodic trajectories through iterative methods (see \S \ref{Sec: DiffCorr}). Numerically integrating for the coefficients of $\STM$ along with $\state$ yields 42 equations of motion. 

\subsubsection{Rotation Matrix} \label{Sec: RotMat}
The frame of reference natural to the dynamics of the CR3BP, the synodic frame, co-rotates with the massive objects in the system. However, to develop physical intuition it is helpful to project solutions from the synodic frame into an inertial frame of reference. This transformation also aids in quantitative comparison to analytic two-body solutions that are calculated in an inertial frame. Given a synodic state vector $^R\state$, the corresponding inertial state vector, $^I\state$, is calculated as,

\begin{equation}
    ^I\state = 
    \begin{bmatrix}
        ^I\mathbf{Q}_R(f)  & 0_{3 \times 3} \\
        \partial_f {^I\mathbf{Q}_R(f)} & ^I\mathbf{Q}_R(f) \\
    \end{bmatrix} 
    {^R{\state}},
    \label{Eq: full_rotmat}
\end{equation}

where, 
\begin{equation}
    ^I\mathbf{Q}_R(f) = \begin{bmatrix}
        \cos{f} & -\sin{f} & 0 \\
        \sin{f} & \cos{f} & 0 \\
        0 & 0 &  1 \\
    \end{bmatrix}
\end{equation}

is a rotation matrix about $\hat{z}$ through angle $f$, the true anomaly of the binary. 

\subsection{Dynamical Systems Theory} \label{Sec: DST}

The dynamical systems theory perspective of the CR3BP was formulated in the late 19\textsuperscript{th} Century by \citet{Poincare1892}. Increased computational capabilities and emerging numerical techniques in the past century have enabled rich insights into the nonlinear system's solution space \citep{Stromgren1922, Szeb1967, BosanacHowellFischbach2015}. In this investigation, we are motivated by recent exoplanet observations to identify the families of solutions that most resemble circumbinary orbital behavior and assess their stability properties by means of dynamical systems theory techniques. For context, a similar analysis of periodic trajectories in the CR3BP provides the foundational dynamical understanding of the Lagrange point `Halo' orbits leveraged as baseline trajectories for spaceflight applications \citep{Breakwell1979, Howell1984}.

\subsubsection{Periodic Solutions} \label{Sec: PerSol}

For a circumbinary exoplanet to persist long enough to be observed, its motion must be bounded to a region around the binary. Compared to the 2BP, sensitivities to initial conditions in the CR3BP make it more difficult to classify regions of state space as bounded motion\footnote{The 2-body problem permits a general region of bounded motion, i.e. $e < 1$.}. In a dynamical system, there are three classifications of bounded motion: fixed-points, limit cycles (periodic orbits) \footnote{Any bounded solution to the 2BP, i.e., ellipse, circle, is a limit cycle.}, and quasi-periodic orbits. Of the bounded motion types, we investigate limit cycles as they provide a broad representative solution space to CBPs and analog the solutions to 2BP bounded motion. 

A limit cycle, $\mathbf{\Gamma}$, is a parameterized closed curve in state space representing a particular solution to a set of ODEs such that the trajectory repeats at some regular time interval, the period, $T$,
\begin{equation}
    \mathbf{\Gamma}(t) : \mathbb{R} \rightarrow \mathbb{R}^6 \ | \ \mathbf{\Gamma}(0) = \mathbf{\Gamma}(T)
\end{equation}

Limit cycles (which we interchangeably call periodic trajectories) in the CR3BP exist as continuous families of solutions and generally propagate their stability properties into nearby solutions. Note that limit cycles are the CR3BP's fundamental, non-static, bounded solutions. 

\subsubsection{Floquet theory} \label{Sec: FloquetTheory}

\begin{figure}[t]
    \centering
    \includegraphics[width = \linewidth]{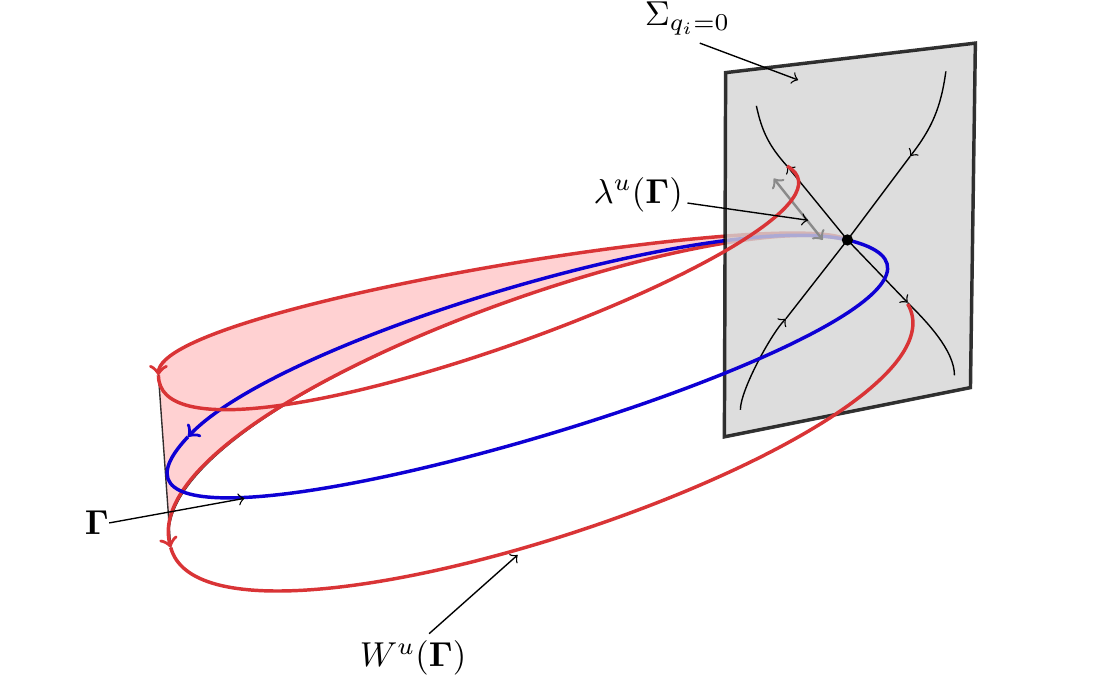}
    \caption{Illustration of Floquet theory (\S\ref{Sec: FloquetTheory}) in a 3-dimensional system. The blue periodic trajectory, $\mathbf{\Gamma}$ generates a fixed-point on the Poincaré section, $\Sigma_{q_i = 0}$. Trajectories initially perturbed in the positive and negative directions (red) of the unstable manifold, $W^u(\mathbf{\Gamma})$, are displaced by the magnitude of the Floquet multiplier $\lambda^u(\mathbf{\Gamma})$ after one period. Note that the trajectories are time-symmetric and therefore perfectly reversible.}
    \label{fig:floquet}
\end{figure}

Recall from \S\ref{Sec: STM} that the linear responses to initial perturbations propagated along a reference trajectory are carried in the STM, $\STM$. Floquet theory asserts that the time-varying coefficients of the State Transition Matrix, $\Phi_{ij}(t)$, along a periodic trajectory may be decomposed into two matrices with periodic coefficients, $\mathbf{F}(t)$ and the exponentiated diagonal matrix $\mathbf{\Omega}$ \citep{Floquet1883, Bosanac2016}. After one period, $\mathbf{F}(T) = \mathbf{F}(0)$ and the resulting STM is 
\begin{equation} \label{Eq: mondromy}
    \mathbf{M} := \mathbf{\Phi}(T, 0) = \mathbf{F}(T)e^{\mathbf{\Omega} T}\mathbf{F}(0)^{-1}
\end{equation}
where $\mathbf{M}$ is the monodromy matrix.  The eigenvalues of $\mathbf{M}$
\begin{equation} \label{Eq: FloquetMult}
    \lambda_i = \pm e^{\Omega_i T}
\end{equation}
are the \textit{Floquet multipliers}, and describe the stretching and rotation of perturbations in the associated eigendirection after one period. By considering the periodic trajectories in state space as fixed points on a Poinacré section (Figure \ref{fig:floquet}), we are able to extract dynamical information about the surrounding solution space. Since the CR3BP is a Hamiltonian system, the six Floquet multipliers of limit cycles come in pairs, $\lambda_a, \lambda_b$, such that
\begin{equation}
    \lambda_a = \frac{1}{\lambda_b}.
\end{equation}
This allows for a decomposition of the 6-D state space into three unique subspaces surrounding a limit cycle. If a limit cycle is a member of a continuous family, one pair of Floquet multipliers is equal to unity, $\lambda_a = \lambda_b = 1$. The remaining two pairs will often be either real ($\lambda_a, \lambda_b \in \mathbb{R}$), or complex on the unit circle ($\lambda_a, \lambda_b \in \mathbb{C} ; |\lambda_a| = |\lambda_b| = 1 $). If $|\lambda| < 1$, a perturbation in the corresponding eigendirection results in exponential decay towards the limit cycle along a \textit{stable manifold}. If $|\lambda| > 1$, a perturbation along the associated eigendirection will exponentially grow from the limit cycle along an \textit{unstable manifold}. In the case $|\lambda| = 1$, the perturbation results in bounded motion near the limit cycle. 

\subsubsection{Stability Indices} \label{Sec: StabInd}

In this investigation, we will use two indices derived from Floquet multipliers to assess the dynamical properties of limit cycles in the CR3BP. The first index, the Poincaré exponent,
\begin{equation} \label{Eq: PCExp_def}
    \Omega_k = \frac{1}{T}(\text{Log}|\lambda^k| + i\text{Arg}(\lambda^k)), \ k \in [1,..., 6],
\end{equation}
describes the $e$-folds of growth or decay of an initial perturbation in the vicinity of a periodic trajectory along the Floquet multiplier's eigendirection after one period. This index is interpreted similarly to a Lyapunov exponent. However, unlike Lyapunov exponents, Poincaré exponents are measured from the monodromy matrix of a single periodic trajectory, rather than the divergence between two reference trajectories. If $\lambda^k \in \mathbb{C}$, the Poincare exponent also describes the angular frequency of oscillation in state space about the reference trajectory. Note, the real part of the Poinacré exponent is zero for any eigenvalues located on the complex unit circle. 

The second index, 
\begin{equation} \label{Eq: Nu_def}
    \nu_i := \frac{1}{2}(\lambda^i_a + \lambda^i_b), \ i \in [1, 2, 3],
\end{equation}
reduces each pair of Floquet multipliers into one real-valued parameter that becomes useful for detecting bifurcations in the solution space. If a non-unity $\nu$ parameter instantaneously equals unity, $\nu_i = \pm 1$, a bifurcation has occurred. The behavior in the limits $\nu_i \rightarrow \pm 1$ indicates the bifurcation type. Bifurcation classification may also be understood through Floquet multipliers transitioning on the complex plane and unit circle. See \citet{Bosanac2016, Gupta2020, Spreen2021} for a detailed explanation of bifurcation classification in the CR3BP. 

\section{Application of Dynamical Systems Theory to the Copenhagen Problem} \label{Sec: CP}

While \citet{Poincare1892}, developed the first periodic trajectories in the CR3BP using perturbative methods at small mass ratios, \citet{Stromgren1938}'s staff of computers at Copenhagen Observatory implemented numerical integration to produce the first limit cycles of an equal-mass CR3BP, the Copenhagen Problem \citep{Szeb1967}. In this section, we reexamine the periodic circumbinary solutions to the Copenhagen problem with additional analysis of the stability properties and dynamical behavior described in \S\ref{sec:dynamical_background}. 

\subsection{Periodic Family Initialization}
We are interested in applying dynamical systems theory to better understand the orbital dynamics of a planet orbiting a circular binary star. The fundamental solutions of circumbinary motion are represented by concentric, near-circular trajectories around each primary mass \citep{Stromgren1922, Szeb1967, BosanacHowellFischbach2015, Bromley2015, BromleyKenyon2021}. In a reference frame rotating with the binary stars, these form a continuous family of periodic trajectories. The computation of these periodic solutions is a boundary condition problem. While a limit cycle contains a continuous set of state vectors, it can also be uniquely identified by one state vector within the set. Following convention, we use the perpendicular $+\hat{x}$ crossing state vector, $\state^* = [x_0^*, 0, 0, 0, \dot{y}_0^*, 0]^T$, as the unique identifier of periodic orbits and their initial conditions for propagation. 

To initialize the differential corrections algorithm (described below) for solving the circumbinary limit cycles in the CR3BP, a nearly-periodic first guess is needed. To first order, multi-pole expansion of the binary's potential equates to two-body EOM. Therefore, a circular Keplerian solution should generate a reasonable first guess to solve for limit cycles far from the binary. We also observe that the autonomous nature of the CR3BP results in the initial true anomaly of the binary having no impact on solutions in the rotating frame. With no loss of generality, we choose $f = 0, ^I{\dot{x}} = 0, ^I{y} = 0$. Noting that non-zero inclination is generally not commensurate with periodic motion in the rotating frame, we take $z = \dot{z} = 0$. Implementing the assumptions above and assuming a Keplerian circular orbit, we generate an initial guess for circumbinary periodic motion in synodic frame coordinates using Equation \ref{Eq: full_rotmat} along $+\hat{x}$, 

\begin{equation}
    \state^* \approx [x, 0, 0, 0, -x \pm x^{-1/2}, 0]^T
\end{equation}

The $\pm x^{-1/2}$ arises from the circular Keplerian approximation for prograde and retrograde motion when viewed in the inertial frame.\footnote{As the orbital period trends upward with increased distance from the barycenter, all circumbinary motion is retrograde when viewed in the synodic frame. We, therefore, refer to \textit{retrograde} and \textit{prograde} exclusively as the motion would appear in the inertial frame.} Consequently, we expect two families of circumbinary limit cycles to exist in the CR3BP. These families were demonstrated to exist by Poincaré through analytic continuation \citep{Szeb1967}. We solve for the first periodic orbits at five binary orbital separations along $\hat{x}$, $x_0 = 5 \ a_\mathrm{bin}$,  which satisfies the assumptions of the estimated initial conditions and encompasses all innermost transiting exoplanets in circumbinary systems \citep{Kostov2021}. 

\subsubsection{Differential Corrections Numerical Algorithm} \label{Sec: DiffCorr}

\begin{figure*}[t]
    \gridline{
                \leftfig{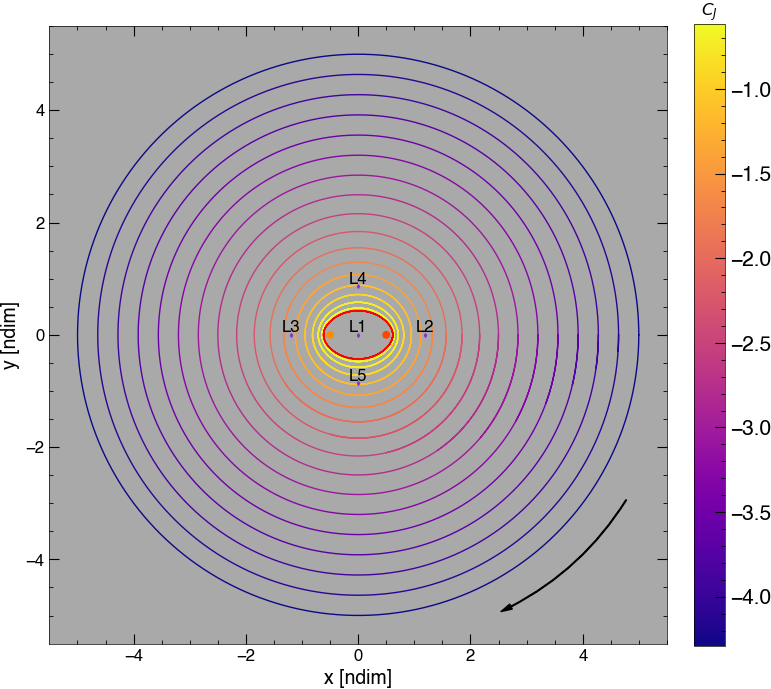}{0.5\textwidth}{(a) Synodic Frame}
                \rightfig{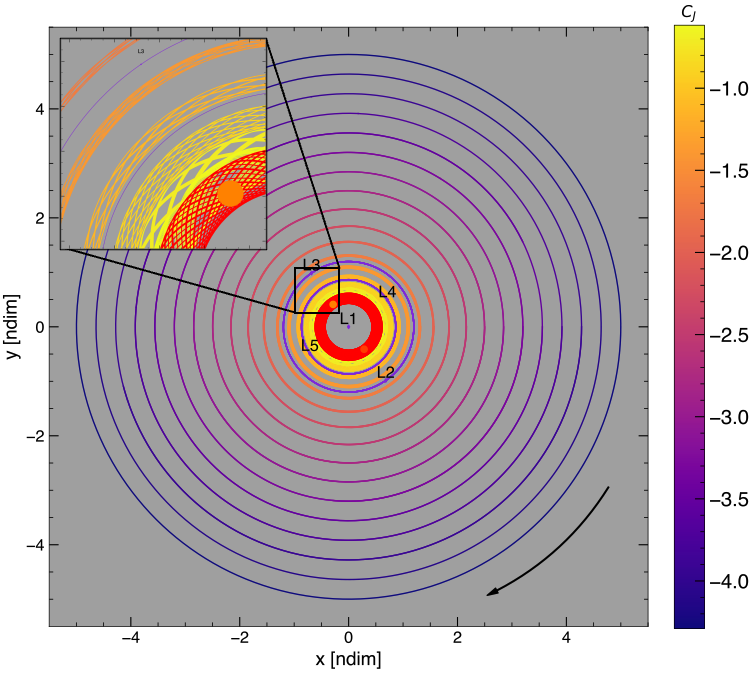}{0.5\textwidth}{(b) Inertial Frame}
    }
    \caption{A sparse representation of the retrograde family of periodic orbital solutions around an equal-mass binary in the synodic (left) and inertial (right) frames. The arrow denotes the motion of the test particle trajectories in each reference frame. The binary is fixed in the synodic frame and rotates counter-clockwise in the inertial frame. The five Lagrange points (fixed points in the synodic frame) are shown. The trajectories far from the binary exhibit near-circular orbits in the synodic frame and flatten along the $\hat{y}$ as the family approaches the binary. The innermost stable orbit (red, see \S\ref{Sec: BifDet}) is interior to all Lagrange points.  It has an elliptical shape in the synodic frame, whereas this same orbit forms an intricate quasi-periodic geometry in the inertial frame.}
    \label{fig:ret_fam}
\end{figure*}

Differential targeting methods implement a gradient-based approach for identifying the desired solution. In this investigation, we implement a simple targeting algorithm, a single shooting method, to compute the initial conditions of circumbinary periodic orbits \citep{Pavlak2010, Bosanac2016, Spreen2021}. The single shooting algorithm is developed with a design vector, $\Xd$, constraint function $\Fconst$, and design Jacobian, $\frac{\partial \Fconst}{\partial \Xd}$. The design vector, $\Xd$, contains $n$ components and is updated at each iteration. The constraint function, $\Fconst$ maps the design variables, $\Xd$ to an $m$ dimensional vector function expressing $m$ scalar constraint equations. This mapping occurs through numerical propagation of the CR3BP equations of motion and measurement of the error between the objective state $\state_0$ and final propagated state, $\state_f$, produced by the design vector, $\Xd$.

The algorithm solves the design variables $\state^*_d$ through iteration such that $\mathbf{G}(\state^*_d) = \bar{0}$. Taking the first order Taylor expansion around $\mathbf{G}(\state^*_d) = \bar{0}$, 
\begin{equation}
    \Fconst \approx \mathbf{G}(\state^*_d) + \frac{\partial \Fconst)}{\partial \Xd} (\Xd - \state^*_d)
\end{equation}
and recasting, $\state^*_d \rightarrow \state_d^{j+1}$, $\Xd \rightarrow \state_d^{j}$, as an update equation,

\begin{equation} 
    \mathbf{G}(\state_d^{j}) +  \frac{\partial \mathbf{G}(\state_d^{j})}{\partial \state_d^{j}}(\state_d^{j+1} - \state_d^{j}) = \bar{0}
\end{equation}
we recover an equation for an updated design vector, $\state_d^{j+1}$. If the number of design variables exceeds the number of constraints, $n>m$, $\frac{\partial \mathbf{G}(\state_d^{j})}{\partial \state_d^{j}}$ is a rectangular matrix with no inverse, i.e., singular, and an infinite number of solutions exist. A unique solution may be obtained through taking the minimum norm solution,
\begin{equation} \label{Eq: MinNorm_Update}
    \begin{split}
    & \mathbf{\bar{X}}^{j+1} = \\ &\mathbf{\bar{X}}^{j} - \frac{\partial \mathbf{G}(\state_d^{j})}{\partial \state_d^{j}}^T\left(\frac{\partial \mathbf{G}(\state_d^{j})}{\partial \state_d^{j}} \cdot \frac{\partial \mathbf{G}(\state_d^{j})}{\partial \state_d^{j}}^T\right)^{-1}\mathbf{G}(\state_d^{j}).
    \end{split}
\end{equation}
If $n<m$, the system is over-determined and no solutions exist. Equation \ref{Eq: MinNorm_Update} can be iteratively solved until a set of design variables, $\state^*_d$, is found such that 
\begin{equation*}
    |\mathbf{G}(\state^*_d)| < \epsilon 
\end{equation*}
where $\epsilon$ is the predetermined tolerance of the solution. In this investigation, periodic family initial conditions are found to $\epsilon \leq 1\times10^{-10}$. 

The most challenging aspect of differential corrections methods is often obtaining the Jacobian matrix that maps the linear response of the constraint vector, $\Fconst$ as a function of the design vector, $\Xd$. Fortunately, $\STM$ contains this information for state variables in the design vector, $\Xd$. In the case that propagation time is included in the design vector, $T \in \Xd$, the Jacobian coefficients relating variations in propagation time to the final state vector, $\state_f$ are contained in $\dstate_f$. For detailed descriptions of differential correction algorithms applied in the CR3BP, see \cite{Pavlak2010, Bosanac2016, Spreen2021}.

\begin{figure*}[t]
    \centering
    \gridline{
                \leftfig{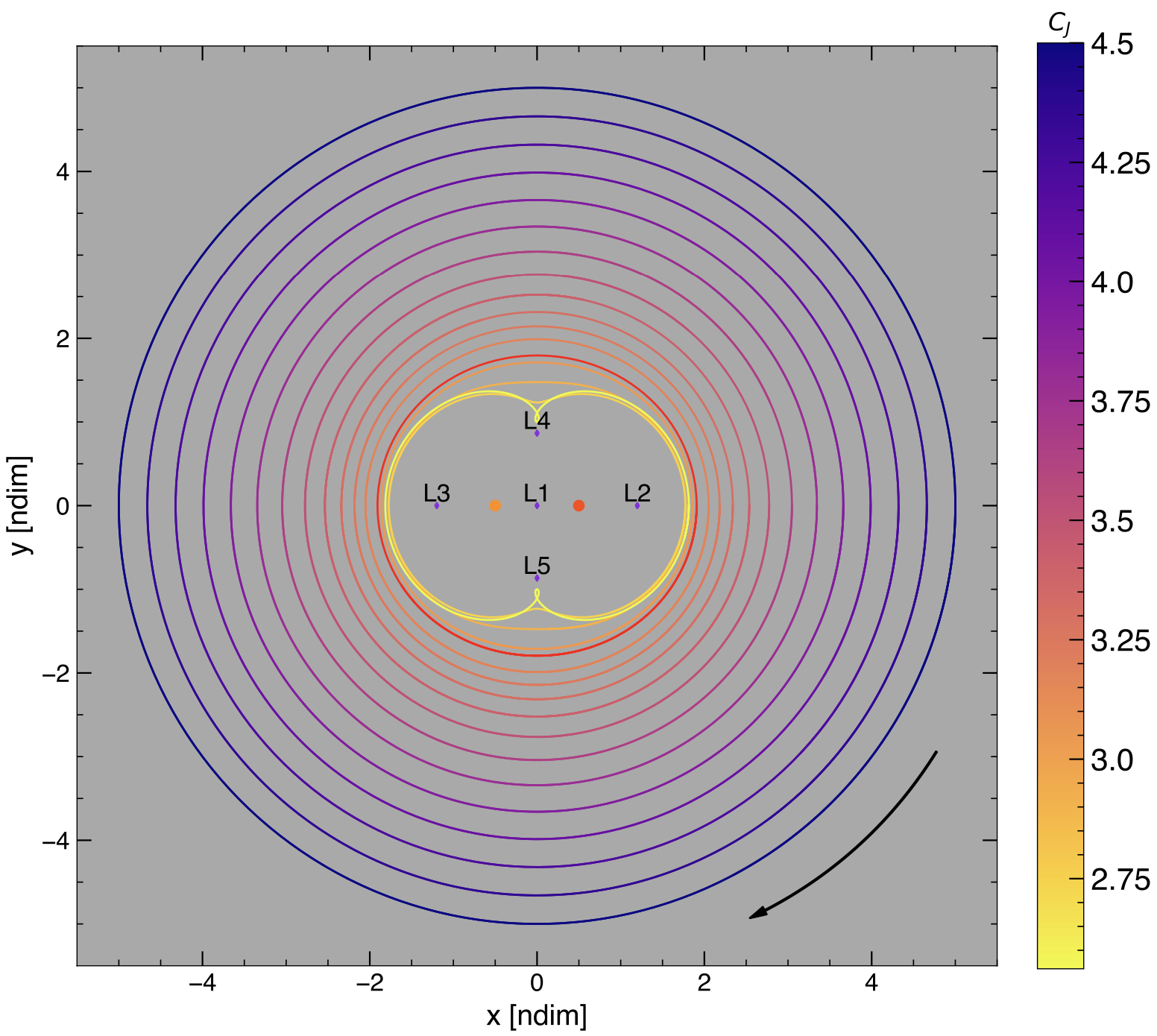}{0.5\textwidth}{(a) Synodic Frame}
                \rightfig{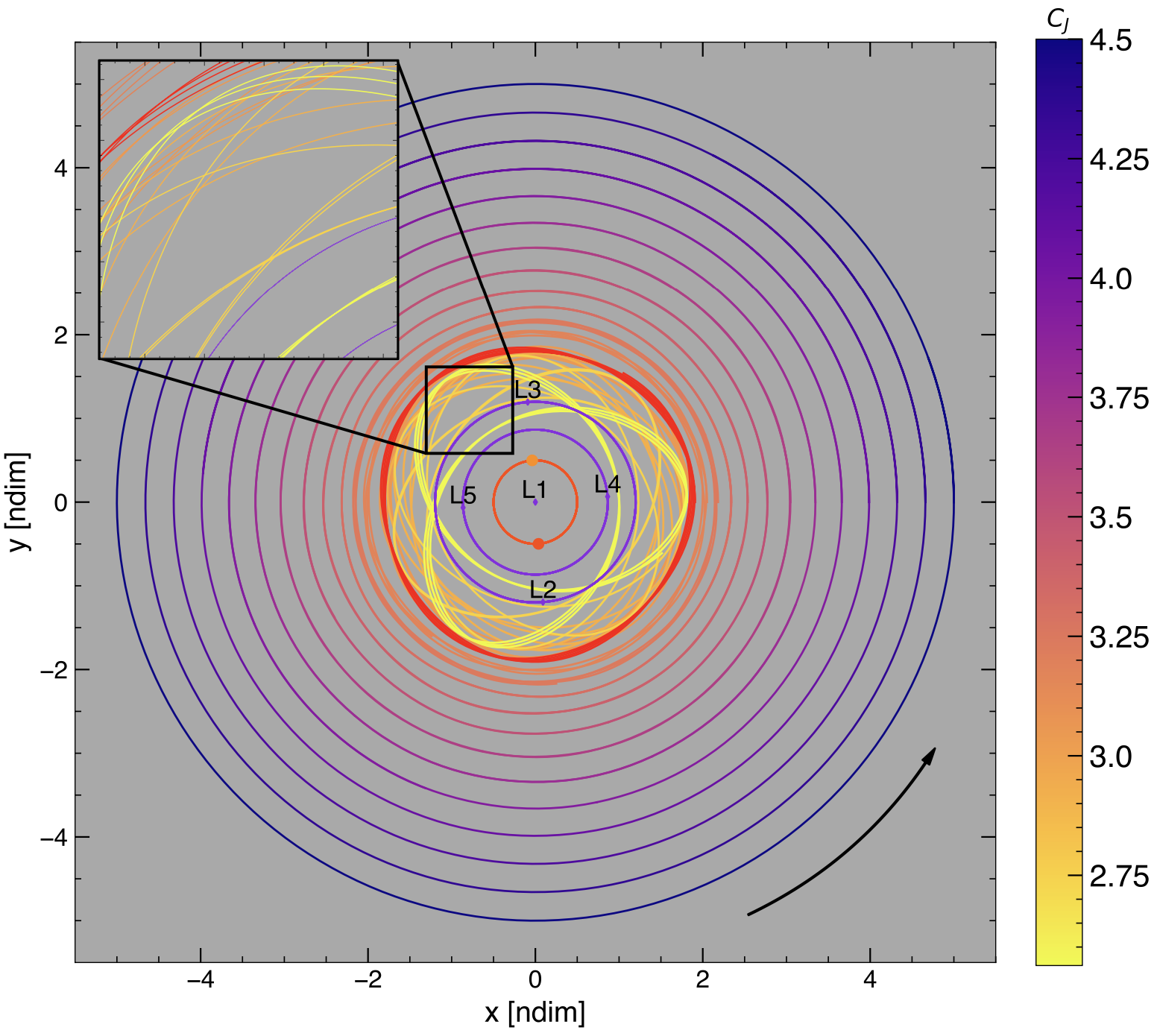}{0.5\textwidth}{(b) Inertial Frame}
    }
    \caption{Same as Figure \ref{fig:ret_fam}, but for prograde periodic orbits. Trajectories far from the binary appear nearly circular, as do trajectories interior to the in-plane bifurcating limit cycle (red). At lower Jacobi constants, the trajectories begin to cusp as they cross $x = 0$ in the synodic frame (left), eventually generating loops near L4 and L5. The inertial frame (right) shows that the cusping orbits rapidly precess about the barycenter, with the precession timescale comparable to the orbital timescale. The in-plane bifurcating orbit (red) does not reveal remarkable geometric features in either frame. }
    \label{fig:pro_fam}
\end{figure*}
\subsubsection{Continuation of Periodic Families} \label{Sec: InitContFam}

A choice of design vector, $\state_d = [x_0, \dot{y}_0, T]$, and constraint vector, $\Fconst = [y_f, \dot{x}_f]$ will solve for solutions that begin and end their trajectories via perpendicular crossing with the synodic x-axis. Per the \textit{Mirror Theorem}, this is a sufficient condition of periodic motion \citep{Roy1995}. The design Jacobian matrix is then given by 
\begin{equation}
    \frac{\partial \mathbf{G}(\state_d)}{\partial \state_d} = \begin{bmatrix}
        \Phi(T, t_0)_{21} & \Phi(T, t_0)_{25} & \dot{y}_f \\
        \Phi(T, t_0)_{41} & \Phi(T, t_0)_{45} & \ddot{x}_f
    \end{bmatrix}, 
\end{equation}
where $\Phi(T, t_0)_{ij}$ is the $i^{th}$ row and $j^{th}$ column component of the propagated STM, $\STM(T, t_0)$. 

To solve for the continuous families, we implement a pseudo-arclength continuation method. Our choice of pseudo-arclength continuation allows a family to be explored when knowledge of its evolution is not known a priori by moving all components of the current periodic design vector, $\state^*_d$, towards an adjacent solution. Once a member of both the retrograde and prograde circumbinary families has been identified, we iteratively apply the pseudo-arclength continuation method to solve family members until a predetermined terminating solution is solved within the family. 

In pseudo-arclength continuation, the next periodic solution is predicted to be in the direction of the design Jacobian matrix null space, $\mathrm{N}\left(\frac{\partial \mathbf{G}(\state_d^*)}{\partial \state_d^*}\right)$. Therefore, the initial guess for the next family member is then given by,

\begin{equation}
    \state_d^{i+1} = \state_d^{*i} \ +\ s\times\mathrm{N}\left(\frac{\partial \mathbf{G}(\state_d^*)}{\partial \state_d^*}\right)
\end{equation}
where $s$ is a scalar value that determines the step size taken in direction of the next solution. For this analysis, we used $s = 5\times10^{-3}$. 

In pseudo-arclength continuation, a choice must be made to stop solving for solutions within the family. We chose to terminate the continuation processes of each family based on the relevance of orbits to observed circumbinary motion. For the prograde family, we stopped continuing the family once the synodic period of the orbits reached $T = 15$ [ndim] as these orbits were well past a near-circular circumbinary geometry of the transiting CBPs \citep{WinnFabrycky2015}. \citet{BosanacHowellFischbach2015} extended the prograde family to $T = 20$ [ndim]. We stopped the continuation of the retrograde family as the $x$-axis crossing location reached a distance of $0.03\, a_\textrm{bin}$ to a primary mass. Interior to $0.03\, a_\textrm{bin}$ of the primary masses, we found the highly nonlinear trajectories are difficult to compute via the simple single shooting algorithm. Our computations generated 1,200--2,500 periodic solutions per family in this investigation.

\subsubsection{Periodic Orbit Geometries in an Equal-Mass Binary} \label{Sec: CPPerGeo}

\begin{figure*}[t]
    \centering
    \includegraphics[width = 0.8\linewidth]{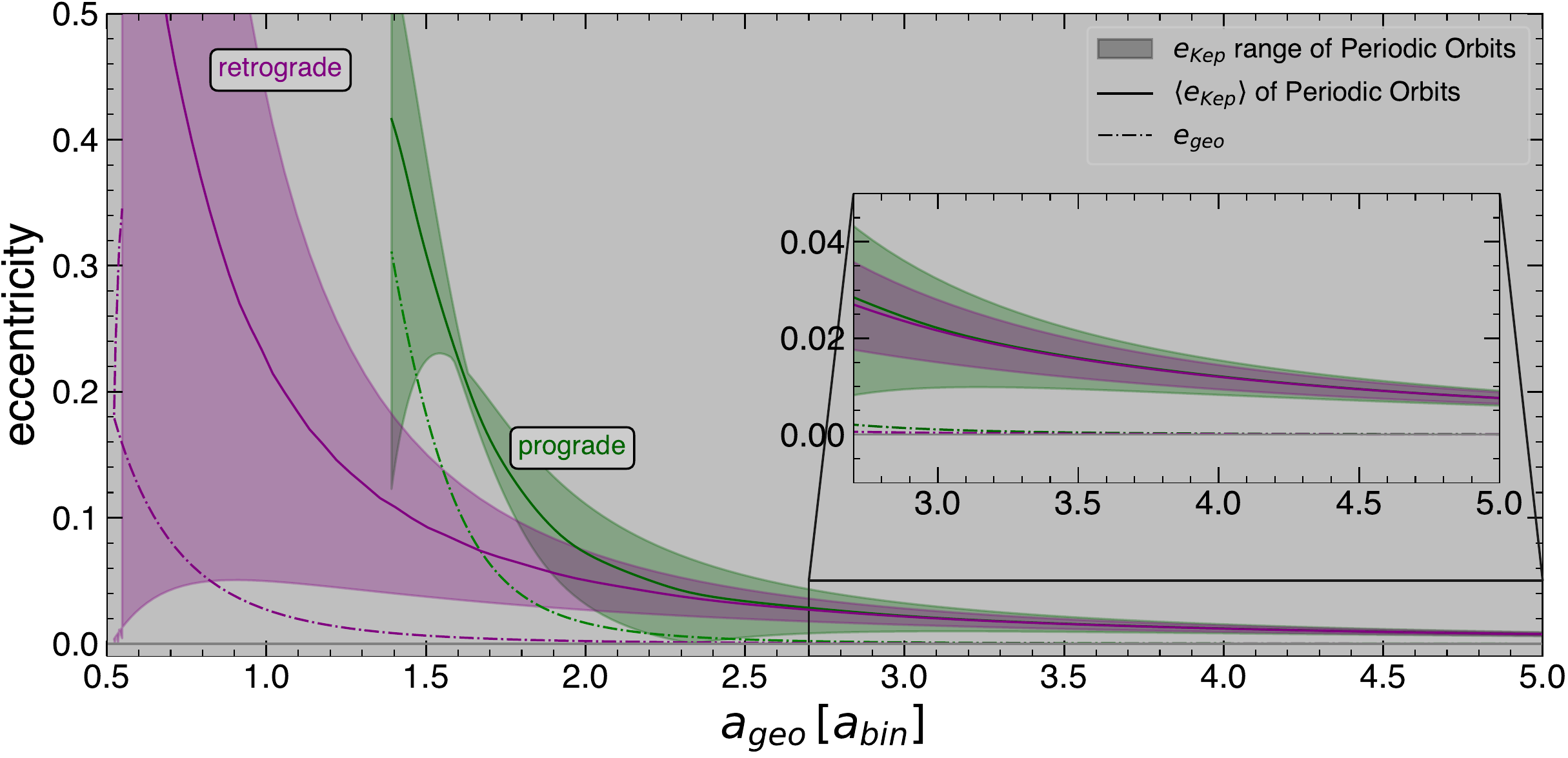}
    \caption{Eccentricity vs. geometric semi-major axis, $a_\mathrm{geo} = (r_a + r_p)/2$, for retrograde (pink) and prograde (blue) families orbiting an equal-mass binary. For each orbit, the Keplerian osculating eccentricities (Eq. \ref{Eq: e_kep}) vary over one orbital period (shaded regions), and a time-averaged Keplerian eccentricity can be computed for each orbit (thick line). Both the geometric (Eq. \ref{Eq: e_geo}, dashed line) and Keplerian eccentricities grow exponentially as trajectories approach the binary. }
    \label{fig: CR3BPvsKep}
\end{figure*}

Initial insight into the planar circumbinary solution space is gained by visualizing the computed families of periodic solutions in the synodic and inertial frames. Figures \ref{fig:ret_fam} and \ref{fig:pro_fam} present a sparse representation of the circumbinary retrograde and prograde periodic solutions respectively, in a circular, equal-mass binary system, i.e. the Copenhagen Problem \citep{Stromgren1922, Stromgren1938, BosanacHowellFischbach2015}.  In each figure, the left panel shows the orbit geometries in the synodic frame, and the right panel shows the orbit geometries in the inertial frame. In both the prograde and retrograde families, the periodic solutions possess near-circular geometry far from the binary but evolve differently as they move toward the barycenter. The retrograde trajectories noticeably flatten along the $x$-axis as the family moves interior to the Lagrange points. Meanwhile, the prograde trajectories also slightly flatten along binary but reach a minimum x-axis crossing of $x_0 = 1.767\,a_\mathrm{bin}$ in the synodic frame. Continuing past this trajectory in the prograde family results in cusp and looping resonant geometries with lower Jacobi constants and longer periods. Each periodic solution in the rotating frame maps to a quasi-periodic bounded orbit in the inertial frame.\footnote{In the chance case that a limit cycle's period is commensurate with the period of the binary, $T_{\mathbf{\Gamma}}:T_{bin} = m:n \ | \ m, n \in \mathbf{N}$, the trajectory will repeat in the inertial frame with period $m\times n$.} In the synodic frame, the retrograde trajectories possess a larger velocity magnitude resulting in a lower Jacobi Constant (see Equation \ref{Eq: CJ}) and a shorter synodic period. However, upon applying a rotation matrix, Equation \ref{Eq: full_rotmat}, the small synodic velocities of the prograde family are `rotated out' and flip the direction of the trajectory. 

In orbital dynamics dominated by a single potential, Keplerian orbital elements carry useful information about the geometry of trajectories. Because conic sections are not solutions to the CR3BP, the geometric eccentricity is distinct and inconsistent with a trajectory's osculating Keplerian eccentricity. Therefore we adopt \citet{SutherlandKratter2019}'s definition of a geometric semi-major axis, $a_\mathrm{geo}$,
\begin{equation}
    a_\mathrm{geo} := (r_a + r_p)/2,
\end{equation}
and eccentricity $e_\textrm{geo}$,
\begin{equation} \label{Eq: e_geo} 
    e_{\mathrm{geo}} := (r_a-r_p)/(r_a + r_p),
\end{equation}
to characterize the typical distance and radial excursion of a bounded circumbinary orbit, where $r_a$ is the apoapsis of the orbit, and $r_p$ is the periapsis of the orbit. Our definition of $e_{geo}$ (Equation \ref{Eq: e_geo}) differs from the equivalently named term presented in \citet{BromleyKenyon2021}'s Equation 11. The value of $a_{geo}$ is also distinct from \citet{LeePeale2006}'s radius of guiding center, $R_0$, which is either assumed to be the Keplerian semi-major axis or solved through FFT techniques \citep{WooLee2020}.

Measuring the osculating Keplerian eccentricity and orbital velocities of guaranteed bounded solutions (periodic orbits) highlights differences between orbits in circumbinary and single-star systems. Past works investigating circumbinary motion with long N-body simulations have evaluated the osculating orbital inclination and argument of periapse induced by the stellar binary and determined precession patterns within sets of circumbinary motion \citep{Doolin2011, Chen2019}. However, the typical osculating eccentricity of circumbinary motion is nuanced since a trajectory's eccentricity directly relies on initial condition formulation and deviations with respect to a `most circular' trajectory \citep{BromleyKenyon2021}. Applying Keplerian circular velocities (\citet{Holman1999, Doolin2011, Quarles2018, Chen2020}) to trajectories at various distances from the binary generates non-zero free eccentricity trajectories and does not guarantee a robust sampling of typical CBP eccentricities \citep{Bromley2015}. 

Our trajectories are periodic by construction and thus guarantee a well-defined $e_\mathrm{geo}$ for each orbit. We compare this to the Keplerian osculating (instantaneous) eccentricity,
\begin{equation} \label{Eq: e_kep}
    e_\mathrm{Kep} = \sqrt{1 - h^2\left(v^2 - \frac{2}{r}\right)}
\end{equation}
where $\vec{h} = \vec{r} \times \vec{v}$. Equation \ref{Eq: e_kep} is derived from the invariant eccentricity vector of the 2BP and is reproducible through applying the vis-viva equation to \citet{Murray2000}'s Equation 2.135. In a single-star planetary system, a trajectory has near-equivalent\footnote{Only in the 2BP, where conic sections are solutions to the EOM, will $e_\mathrm{Kep} = e_\mathrm{geo}$.} values for $e_\mathrm{geo}$ and $e_\mathrm{Kep}$. In the CR3BP, the geometric eccentricity provides a metric to characterize the shape of bounded orbits that is invariant between reference frames. When applied to a circumbinary orbit, the Keplerian eccentricity measures the difference in angular momentum between a circumbinary and circular two-body orbit at the same distance from the barycenter. 

In the top panel of Figure \ref{fig: CR3BPvsKep}, we show computed values of geometric and the range of osculating Keplerian eccentricities in the Copenhagen Problem prograde (blue) and retrograde (pink) periodic families. The geometric eccentricities of periodic solutions grow as they move toward the binary matching the large radial excursions of the periodic trajectories closest to the binary in Figures \ref{fig:ret_fam} and \ref{fig:pro_fam}. The time-averaged osculating Keplerian eccentricity over one synodic period also grows and is consistently larger than the geometric eccentricity in a given orbit. The notable disagreement between measured geometric and ranges of Keplerian eccentricity highlights their disassociation in the CR3BP. The geometric eccentricity is a particularly relevant metric because a number of astrophysical processes depend on a planet's variation in distance from a stellar body. Note that the relations $r_\textrm{(apo, peri)} = a(1\pm e)$ are valid for CBPs only when $a_\textrm{geo}$ and $e_\textrm{geo}$ are used in place of their Keplerian elemental definitions.

Leveraging the Copenhagen Problem retrograde and prograde periodic families as fundamental solutions to circumbinary motion, we consider their agreement with prior analytic predictions about bounded circumbinary trajectories. The forced eccentricity distribution of a thin 3D disk grows exponentially near the binary \citep{Lubow2022}. \citet{Bromley2015} predicted that CBPs reside on \textit{dynamically cool} trajectories that mimic the role of Keplerian circular orbits around a single star. Their analytic calculations suggested these trajectories exist as nested concentric orbits with a minimal forced eccentricity aligned with the binary -- a description matching the periodic orbits in Figures \ref{fig:ret_fam} and \ref{fig:pro_fam}. They also found that choosing 2BP initial conditions -- as compared to their analytically derived nested, dynamically cool orbits -- produced an excess free eccentricity, and as a result more scattering events. By construction, our differential corrections approach to solving for periodic initial conditions generates sets of trajectories in non-crossing (and therefore dynamically cool) configurations. Therefore we posit the families of circumbinary synodic limit cycles as the set of `most circular' orbits that exist around a zero eccentricity binary \citep{LeePeale2006, YoudinKratterKenyon2012, Bromley2015, BromleyKenyon2021}.

\subsection{Bifurcation Detection and Analysis} \label{Sec: BifDet}
Unlike in the 2BP, CR3BP trajectories may possess differences in long-term stability without incurring an appreciable visual difference in the trajectories' geometry. Therefore, we extract information about the dynamics surrounding a periodic solution in the CR3BP via the Floquet theory techniques presented in \S\ref{Sec: FloquetTheory}. Changes to the stability are identified based on the local topology of the solution space and constitute various types of bifurcations \citep{Broucke1969, Perko1996}. In many cases, bifurcations in the CR3BP are generated where two families tangentially connect in the 6-dimensional state space \citep{Bosanac2016, Spreen2021}. Calculating the evolution of these tangent families is out of the scope of this work as (1) we do not expect them to be stable, and (2) they are unlikely to describe the orbital trajectories of observed CBPs. Nevertheless, knowledge of the full solution space of periodic trajectories in the CR3BP is foundational to a rich understanding of the possible dynamics. 

Because perturbations from exact limit cycles are expected in natural planetary systems, our examination of local dynamics around periodic solutions allows us to determine the long-term behavior and hence, stability, of trajectories near these limit cycles. In the CR3BP, Floquet theory provides relevant computation to precisely locate changes in stability along a continuous family of periodic solutions \citep{Floquet1883, BosanacHowellFischbach2015, Bosanac2016}. The linear response--static, oscillatory, or exponential dynamics--of a perturbation to a periodic trajectory is captured by the Floquet multipliers\footnote{eigenvalues of the monodromy matrix, Equation \ref{Eq: mondromy}} (Equation \ref{Eq: FloquetMult}). Consequently, Floquet theory provides precise information about how linear dynamics evolve around limit cycles in any direction of state space.

\begin{figure*}[t]
    \centering
    \gridline{
                \leftfig{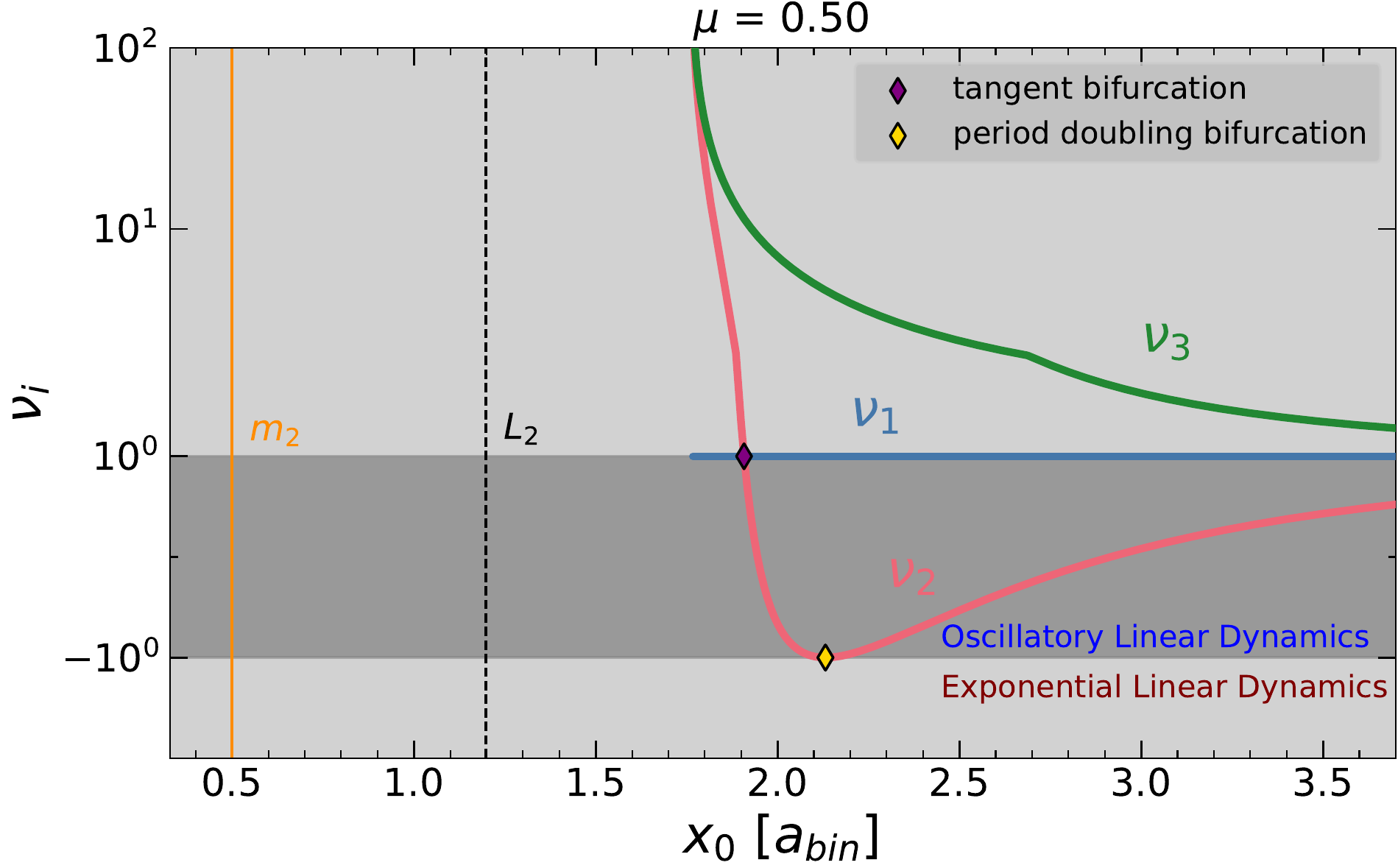}{0.5\textwidth}{(a) Prograde Family}
                \rightfig{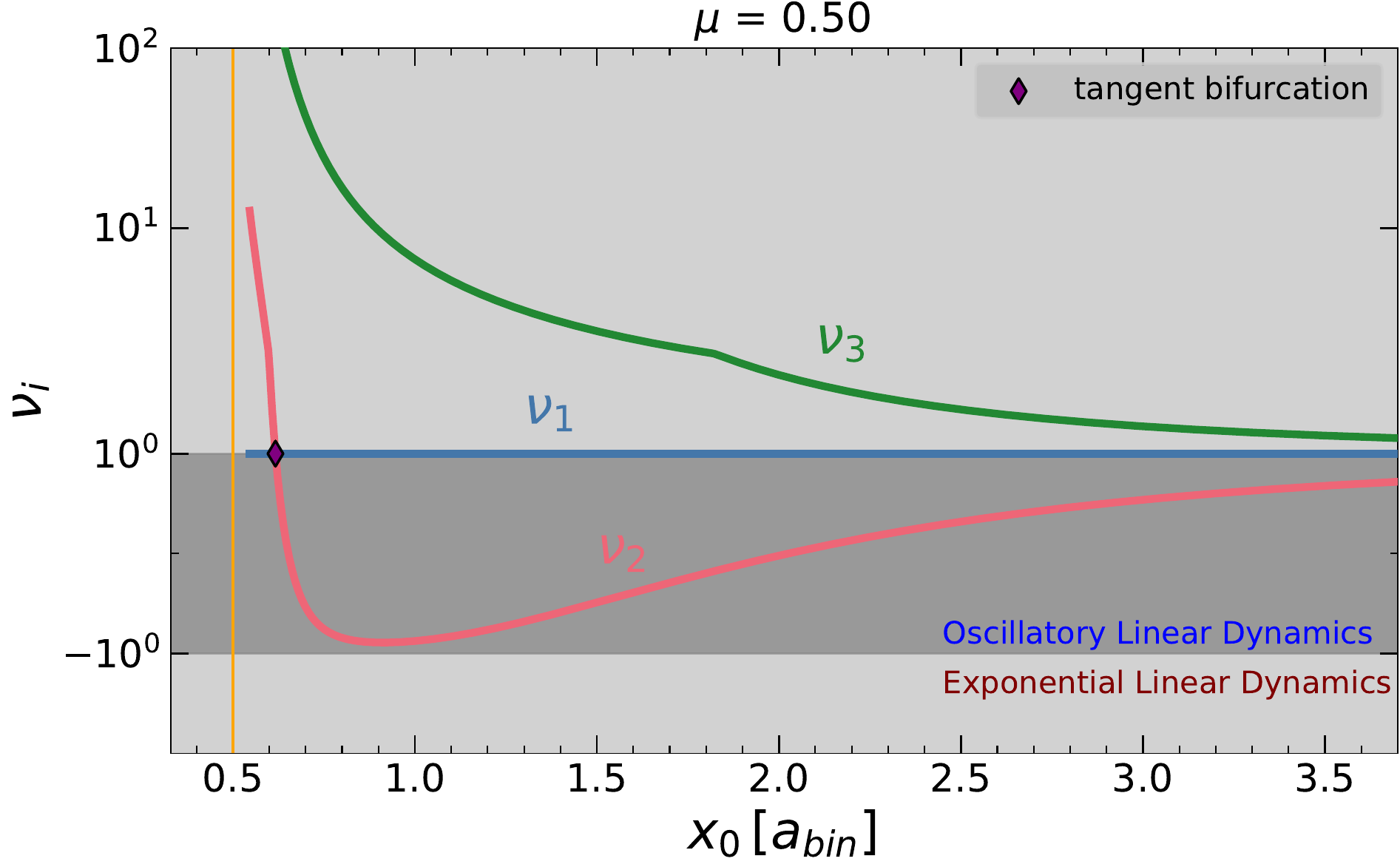}{0.5\textwidth}{(b) Retrograde Family}
    }
    \caption{Bifurcation parameters $\nu_{1}$ (blue), $\nu_{2}$ (red), and $\nu_{3}$ (green) vs. their $x$-axis crossing positions for prograde (left) and retrograde (right) periodic trajectories orbiting an equal-mass binary ($\mu=0.5$).  Regions of oscillatory linear dynamics ($|\nu| < 1$) and unbounded linear dynamics ($|\nu| > 1$) are indicated (note the semi-log scaling of the y-axis). Tangent bifurcations ($\nu = 1$, purple diamonds) occur for $\nu_2$, which is affiliated with the in-plane eigenvector, in both the prograde and retrograde families.  A period-doubling bifurcation ($\nu=-1$, yellow diamond) occurs for $\nu_2$ in the prograde orbital family.  The parameter associated with out-of-plane motion, $\nu_3$, corresponds to exponential dynamics which we examine in more detail in \S\ref{Sec: outofplane}. The parameter $\nu_1$ is unity across each family, consistent with the behavior of a continuous solution space of periodic orbits. See Figures \ref{fig:munu_pro} and \ref{fig:munu_ret} for unequal mass binaries. }
    \label{fig: mu05_nuvx}
\end{figure*}

A bifurcation within a family is denoted by a pair of Floquet multipliers transitioning so perturbations in the associated eigendirection produce a qualitatively different linear dynamical response. In the CR3BP, the monodromy matrix, $\mathbf{M}$, is not symmetric and therefore not guaranteed to have orthogonal eigenvectors. Consequently, bifurcations to one pair of eigenvectors may also implicate a change in stability to perturbations in another non-orthogonal pair. Pairs of eigendirections and their dynamical characteristics are simplified through the bifurcation parameter $\nu$ (Equation \ref{Eq: Nu_def}). Throughout this investigation, $\nu_1$ corresponds to the \textit{unitary} Floquet multiplier pair, which have an in-plane eigendirection tangent to the trajectory. The parameter $\nu_2$, corresponds to the \textit{non-unitary} Floquet multiplier pair with an in-plane eigendirection that varies angle with respect to the tangent eigenvector. As periodic orbits extend further from the binary, the angle between the in-plane eigendirections approaches zero, i.e, $\lim_{x_0 \rightarrow \infty} \hat{e}_i \cdot \hat{e}_j = 0$, where $\hat{e}_i, \hat{e}_j$ are non-paired planar monodromy matrix eigenvectors. The final parameter, $\nu_3$ corresponds to non-unitary Floquet multipliers with an out-of-plane eigendirection, i.e., $z,\dot{z}$ perturbations. As described in \S\ref{Sec: StabInd}, bifurcations are delineated by $\nu_i$ instantaneously equaling $\pm1$.

Figure \ref{fig: mu05_nuvx} shows the evolution of the bifurcation parameter across the prograde and retrograde periodic families in the Copenhagen Problem. By definition, $\nu_1$ remains at unity across the continuous families. As the families extend further from the binary, each $\nu$ parameter approaches unity. The occurrence of three unitary $\nu$ values results in any perturbation leading to a new limit cycle, similar to the 2BP, which the CR3BP approaches as $x_0 \rightarrow \infty$. Values of $\nu_2=-1$ correspond to a period-doubling bifurcation, whereas values of $\nu_2=1$ correspond to tangent bifurcations \citep{Gupta2020}.  For an equal-mass binary, the period-doubling bifurcation only occurs in the prograde direction. See \S\ref{Sec: allmu_orbgeo} for examples at other mass ratios. In both the retrograde and prograde families for $\mu=0.5$, a tangent bifurcation occurs in $\nu_2$, altering the effect of in-plane perturbations from oscillatory ($\nu_2 < 1$) to exponential ($\nu_2 > 1$) with respect to the reference limit cycle. The components of the eigenvectors interior to tangent bifurcations are in each planar direction $x, y, \dot{x}, \dot{y}$. Thus, any in-plane perturbations will grow exponentially in the state space surrounding limit cycles interior to the tangent bifurcation, resulting in long-term instability of the test particle. 

The final pair of Floquet multipliers, characterized by $\nu_3$, have out-of-plane eigenvectors, orthogonal to those of the in-plane pairs, $\nu_1$, $\nu_2$. Perturbations in the out-of-plane direction are characterized by exponential linear dynamics ($\nu_3 > 1$) for all values of $x_0$ we explored, in both the prograde and retrograde families. Often, exponential dynamics will lead to long-term instability around a limit cycle. However, since the $\nu_3$ eigendirections are non-orthogonal, the nonlinear stable and unstable manifolds intersect and bounded motion may still exist under certain conditions. In the retrograde and prograde families, the unstable manifold $W^u(\mathbf{\Gamma})$, is more activated when $\textrm{sign}(z) = \textrm{sign}(\dot{z})$, contributing to exponential growth from the limit cycle. The stable manifold, $W^s(\mathbf{\Gamma})$, is more activated when $\textrm{sign}(z) = -\textrm{sign}(\dot{z})$, contributing to exponential decay towards the limit cycle. This description of growth and decay based on the signs of $z/\dot{z}$ also describes aspects of simple harmonic motion. We show in \S\ref{Sec: UnsBehavior} with numerical simulations that perturbations along the unstable eigendirection approach oscillatory, bounded behavior on long time-scales, and so the test particle is stable to small out-of-plane perturbations. 

\subsection{Investigating Exponential Dynamical Behavior} \label{Sec: UnsBehavior}

In \S\ref{Sec: BifDet} we found that Floquet theory predicts two occurrences of exponential, and therefore unstable, dynamics for circumbinary motion in the Copenhagen Problem. The first occurrence is present throughout both the retrograde and prograde families and is associated with out-of-plane perturbations. The second occurrence involves in-plane perturbations interior to a tangent bifurcation. Sections \ref{Sec: outofplane} and \ref{Sec: inplanebif} will explore the implications of these predictions of exponential linear dynamics by means of numerical experimentation. 

\subsubsection{Out-of-plane perturbations} \label{Sec: outofplane}

Floquet theory predicts exponential dynamics in $\hat{z}/\hat{\dot{z}}$ given a perturbation out-of-plane to the prograde and retrograde limit cycles. However, this behavior is a prediction relative to the planar limit cycle in state space and does not account for the long-term effects of the intersecting stable and unstable manifolds. A simple analytic treatment of the CR3BP EOM predicts bounded motion in $\hat{z}/\hat{\dot{z}}$ under conditions generally satisfied by the circumbinary periodic solutions of the Copenhagen Problem. The following derivation illustrates how perturbations in $\hat{z}/\hat{\dot{z}}$ to near-circular orbits outside the binary are predicted to induce oscillatory motion in $\hat{z}$ at a frequency determined by the distance from the barycenter. 

Beginning with the $\hat{z}/\hat{\dot{z}}$ CR3BP equation of motion, 
\begin{equation}
    \begin{split}
        &\ddot{z} = -\frac{1-\mu}{r_{31}^3}z -\frac{\mu}{{r_{32}^3}}z \\
    \end{split}
\end{equation}
where $r_{31} = \sqrt{(x + \mu)^2 + y^2 + z^2}$ and  $r_{32} = \sqrt{(x - (1- \mu))^2 + y^2 + z^2}$,
we assume $(x + \mu)^2 + y^2 \approx (x - (1- \mu))^2 + y^2 = r^2$, conditions met by near-circular orbits far outside the binary – keeping in mind neither the prograde nor retrograde families follow this assumption near the binary. Assuming $z<< a_\mathrm{bin} <r$, the equation of motion becomes, 
\begin{equation*}
    \begin{split}
        \ddot{z} &\approx -\frac{z}{(r^2 + z^2)^{3/2}} \\
        & \approx -\frac{z}{r^3}\left(1 + \left(\frac{z}{r}\right)^2\right)^{-3/2} \\
        & \approx -\frac{z}{r^3} - \frac{3}{2}\frac{z^3}{r^4} + O(z^5) \\
    \end{split}
\end{equation*}

Taking $z$ to the first order,

\begin{equation} \label{Eq: zpet}
    z(t) \approx \sqrt{z_0^2 + \left(\frac{\dot{z}_0}{\omega z_0}\right)^2}\cos(\omega t - \phi_0)
\end{equation}
where  $\omega = r^{-3/2}$ and $\phi_0 = \tan^{-1}{(\dot{z}_0/\ \omega z_0)}$. Equation \ref{Eq: zpet} predicts that a small perturbation in $\hat{z}$ to a member of the prograde or retrograde limit cycles is bounded in amplitude and oscillates around $z=0$.

\begin{figure}[t]
    \centering
    \includegraphics[width = \linewidth]{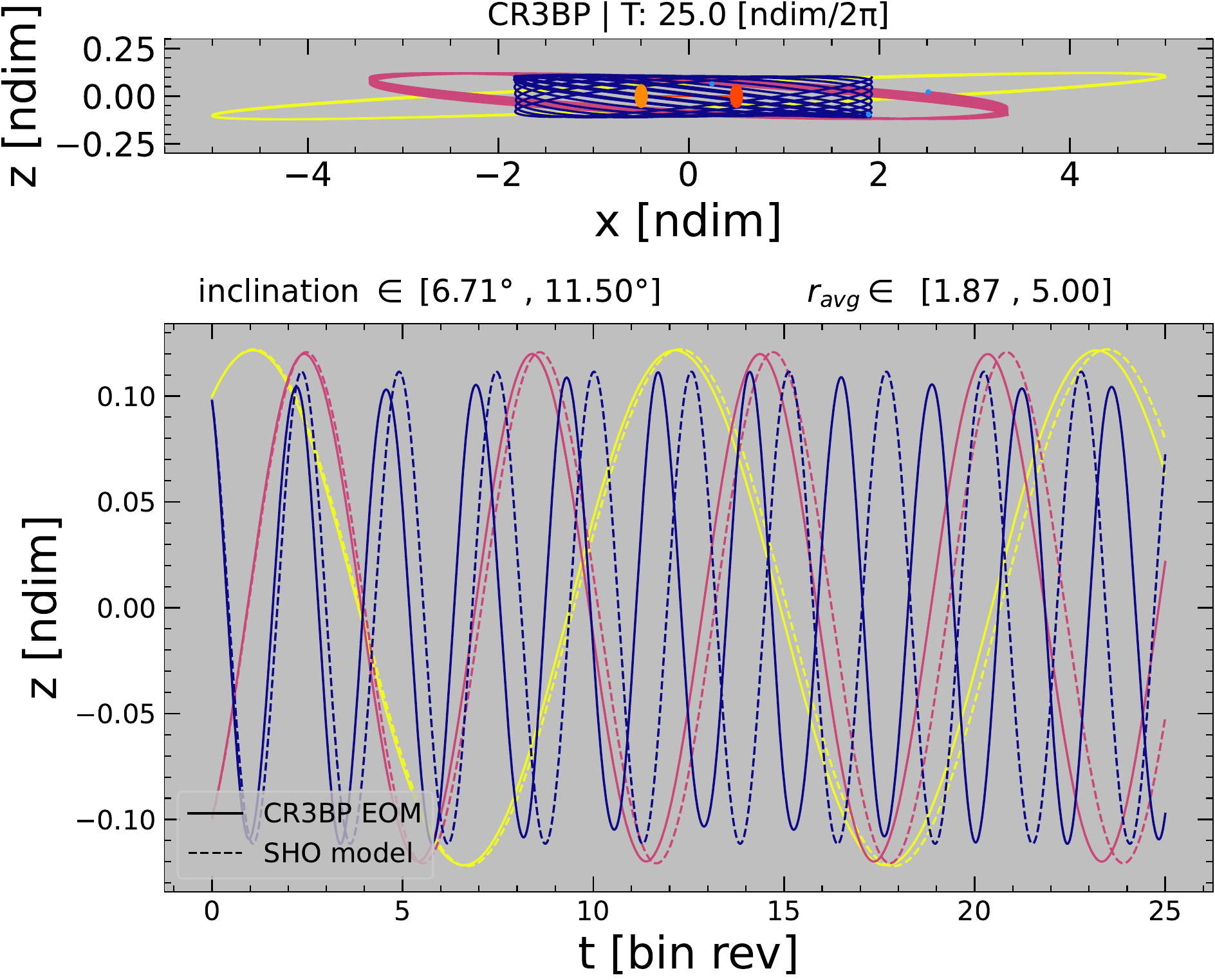}
    \includegraphics[width = \linewidth]{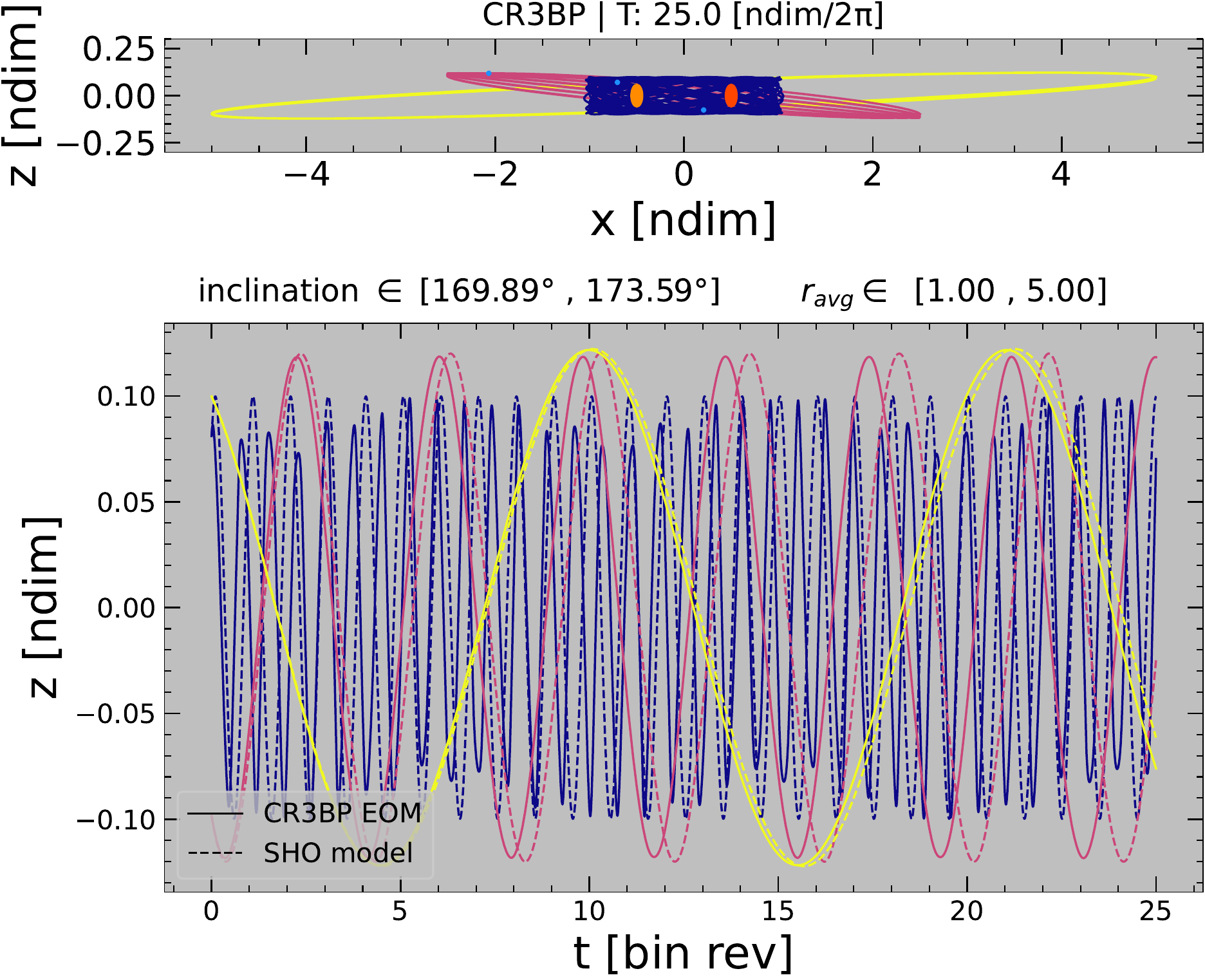}
    \caption{The trajectories of perturbed orbits for prograde (top group) and retrograde (bottom group) limit cycles in the out-of-plane direction, $z/\dot{z}$, along the unstable invariant manifold. Inertial $x-z$ projections (small panels) show the inclined orbits. In the large panels, the propagation of CR3BP EOM over time (solid lines) is compared to an analytic model of SHO derived from the linearized EOM (dashed lines). The analytic model, Equation \ref{Eq: zpet}, generally agrees with the numerical amplitude and angular frequencies of out-of-plane perturbations to circumbinary motion. This suggests the out-of-plane intersecting unstable/stable manifolds can lead to bounded motion in state space on long time scales for sufficiently far distances from the binary.}
     \label{fig: pro_zpet}
\end{figure}

\begin{figure*}[t] 
    \centering
    \gridline{
                \leftfig{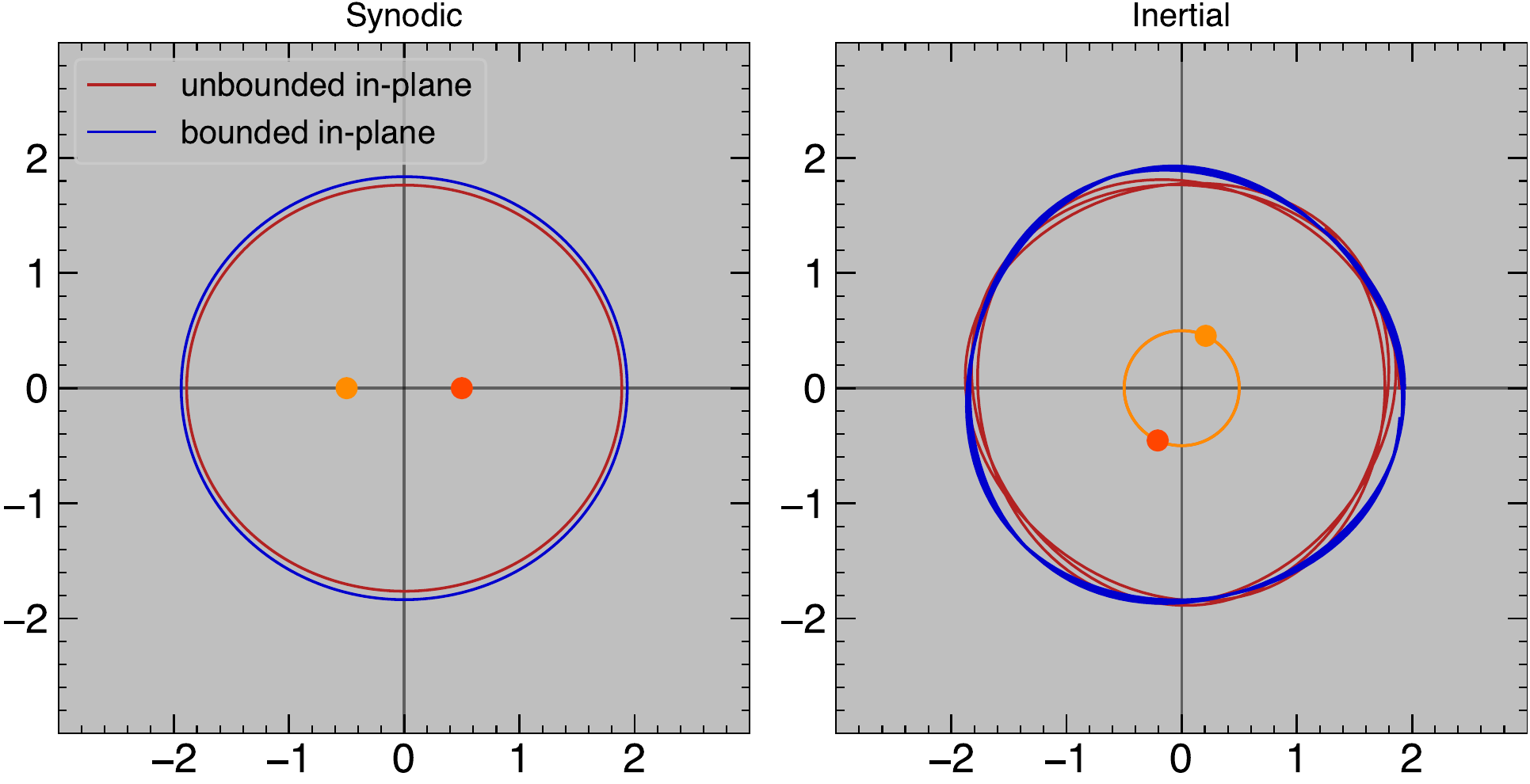}{0.5\textwidth}{(a) Prograde Limit Cycles}
                \rightfig{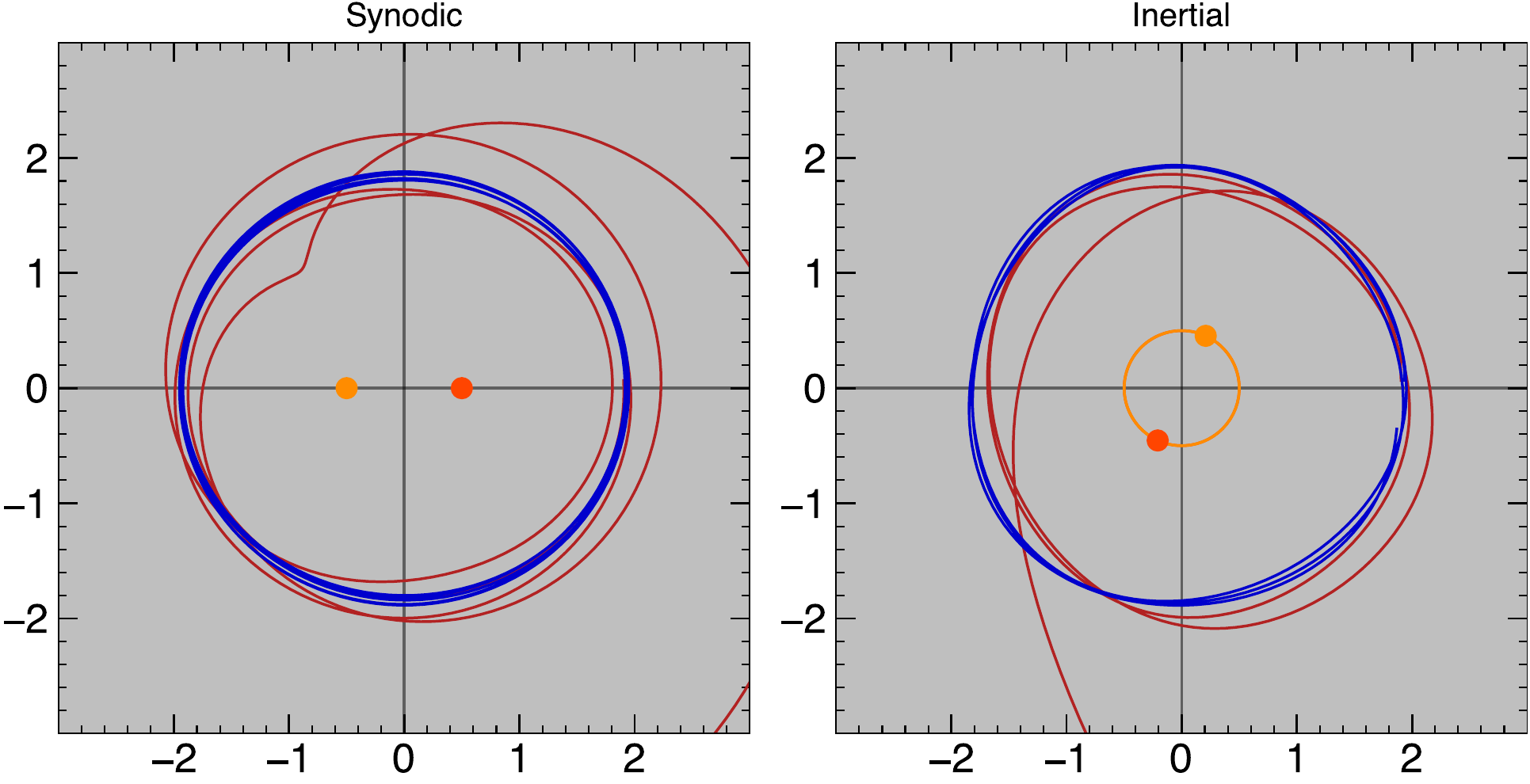}{0.5\textwidth}{(b) Perturbed Prograde Limit Cycles}
    }
    \gridline{
                \leftfig{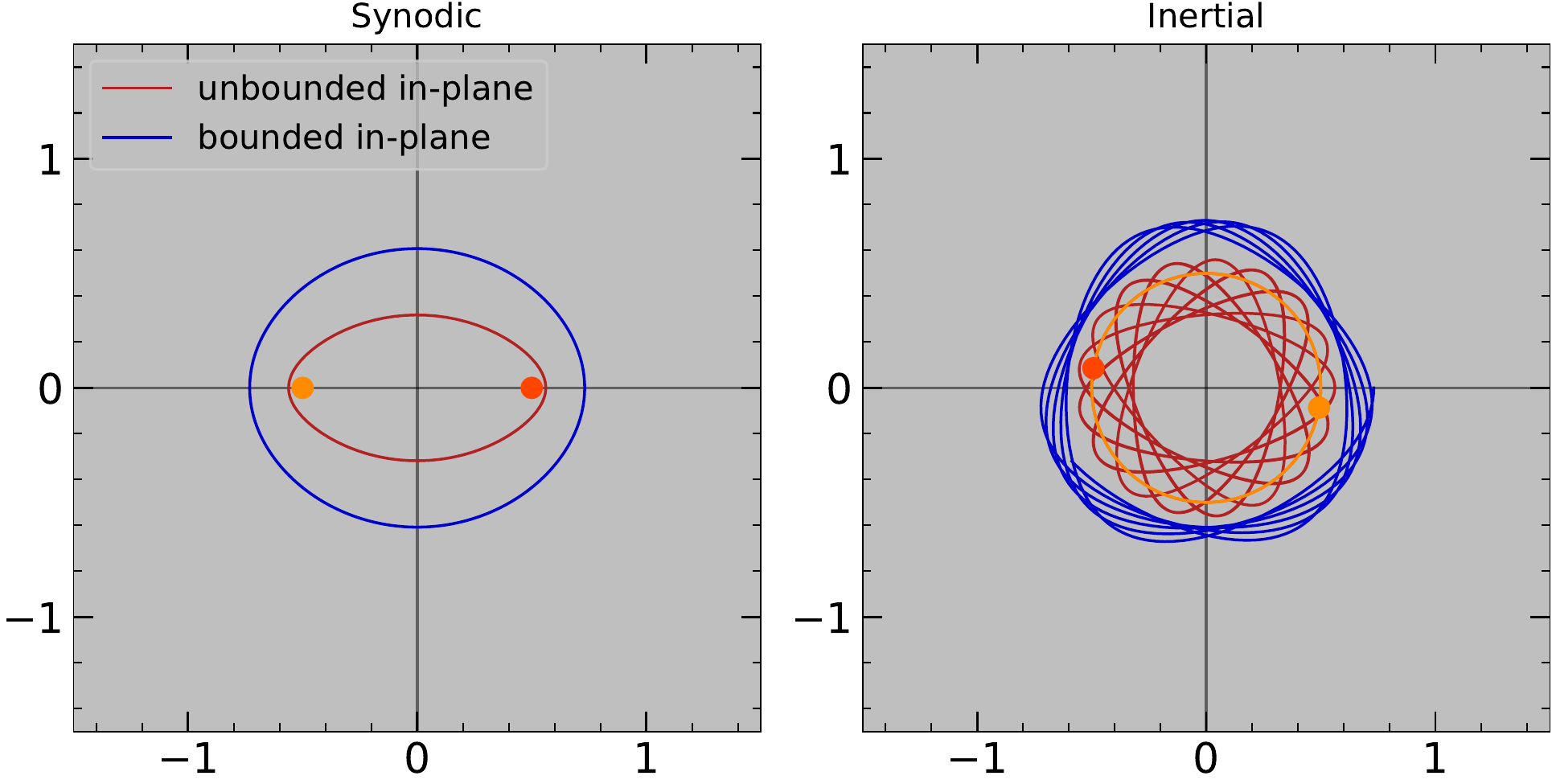}{0.5\textwidth}{(c) Retrograde Limit Cycles}
                \rightfig{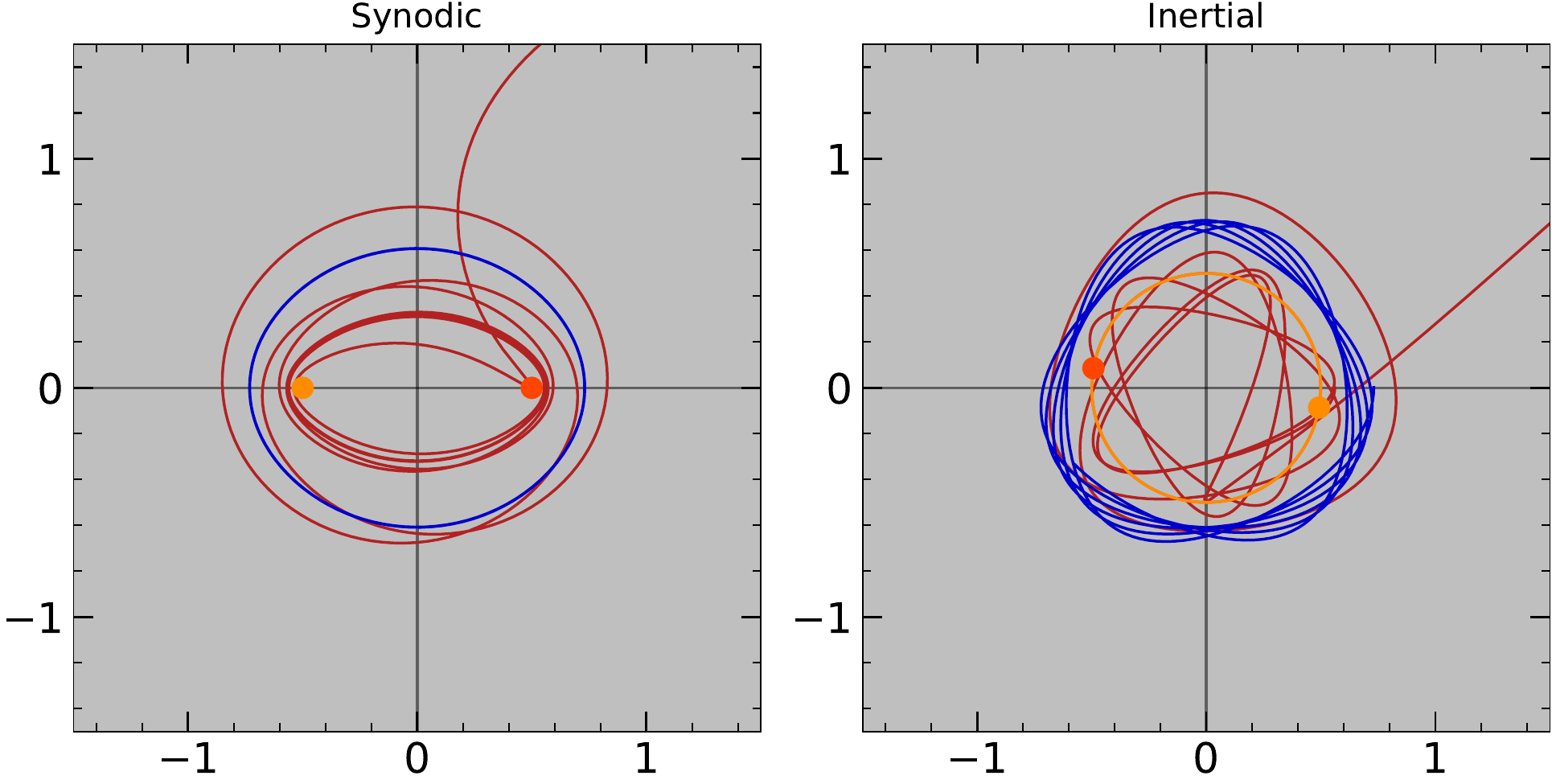}{0.5\textwidth}{(d) Perturbed Retrograde Limit Cycles}
    }
    \caption{In-plane perturbations on geometrically similar orbits produce qualitatively different outcomes depending on whether the orbit is on the bounded (blue) or unbounded (red) side of a stability-changing bifurcation. Top row, left two panels: two periodic trajectories from the prograde family, in both the synodic (left) and inertial (right) frames, with positions and orbits of the stars indicated (orange and red dots).  Top row, right two panels: the orbits are perturbed in the associated eigendirection of the tangent bifurcation. Perturbation of the bounded (blue) orbit leads to quasi-periodic motion around the limit cycle, whereas perturbation of the unbounded (red) orbit leads to an ejection from the binary system. Bottom row: the same as above, but for retrograde orbits that remain bounded (blue) or become unbounded (red) after an in-plane perturbation.  Note the different axes.}
    \label{fig: limit_pert_exp}
\end{figure*}

To test the validity of Equation \ref{Eq: zpet}, we compare the SHO model to numerical CR3BP experiments by perturbing limit cycles out-of-plane at incremental distances from the binary along the unstable manifold directions, Figure \ref{fig: pro_zpet}. The eigenvector direction of the unstable Floquet multiplier provides the appropriate ratio, $z_0:\dot{z}_0$, between out-of-plane components. In the prograde family, the ratio is approximately $9:1$. We determined the magnitude of perturbation by generating inclinations comparable, but larger than the measured inclinations of Kepler CBPs \citep{WinnFabrycky2015}. This guarantees our analysis encompasses a range of inclinations similar to observable CBP systems. 

Figure \ref{fig: pro_zpet} reveals that prograde trajectories perturbed in the $z/\dot{z}$ direction are in close alignment with the predicted maximum amplitude from the SHO model. Most importantly, this agreement even extends close to the in-plane bifurcating orbits in the prograde family. In the retrograde family, we found that perturbations along $z/\dot{z}$ of order $10^{-1}$ were consistent with the bounded motion described by our SHO model for $x_0 > 1\ a_\mathrm{bin}$. For trajectories sufficiently far from the binary, the SHO model also provides a close prediction of the angular frequency of oscillation about the plane.

This numerical experimentation highlights the nuanced role of invariant manifolds as dynamical structures in circumbinary orbital dynamics. Since any physical CBP will have a non-zero inclination, the out-of-plane invariant manifolds will influence the long-term evolution of a planet's trajectory. In a nonlinear dynamical system, the unstable invariant manifold serves as an attracting surface for trajectories in the vicinity and the same Jacobi Constant \citep{Perko1996}. As a consequence, we should expect to find that inclined circumbinary trajectories approximately trace the invariant unstable manifolds of circumbinary limit cycles. 

\subsubsection{In-plane tangent bifurcation} \label{Sec: inplanebif}

The second prediction of instability in circumbinary trajectories arises from tangent bifurcations occurring in the plane of the binary. This feature emerges from the pair of complex-valued Floquet multipliers converging to unity and separating along the real axis \citep{Gupta2020, Boudad2022}.

Figure \ref{fig: limit_pert_exp} shows the effect of perturbations to limit cycles in the bifurcating eigendirection both interior (blue) and exterior (red) to the bifurcation in the synodic and inertial frames. The geometry and solutions of these periodic orbits does not differ substantially. Without perturbation, these mathematical solutions exist as bounded orbits that repeat at regular intervals for all time. 

\begin{figure*}[t]
    \centering
    \includegraphics[width = \linewidth]{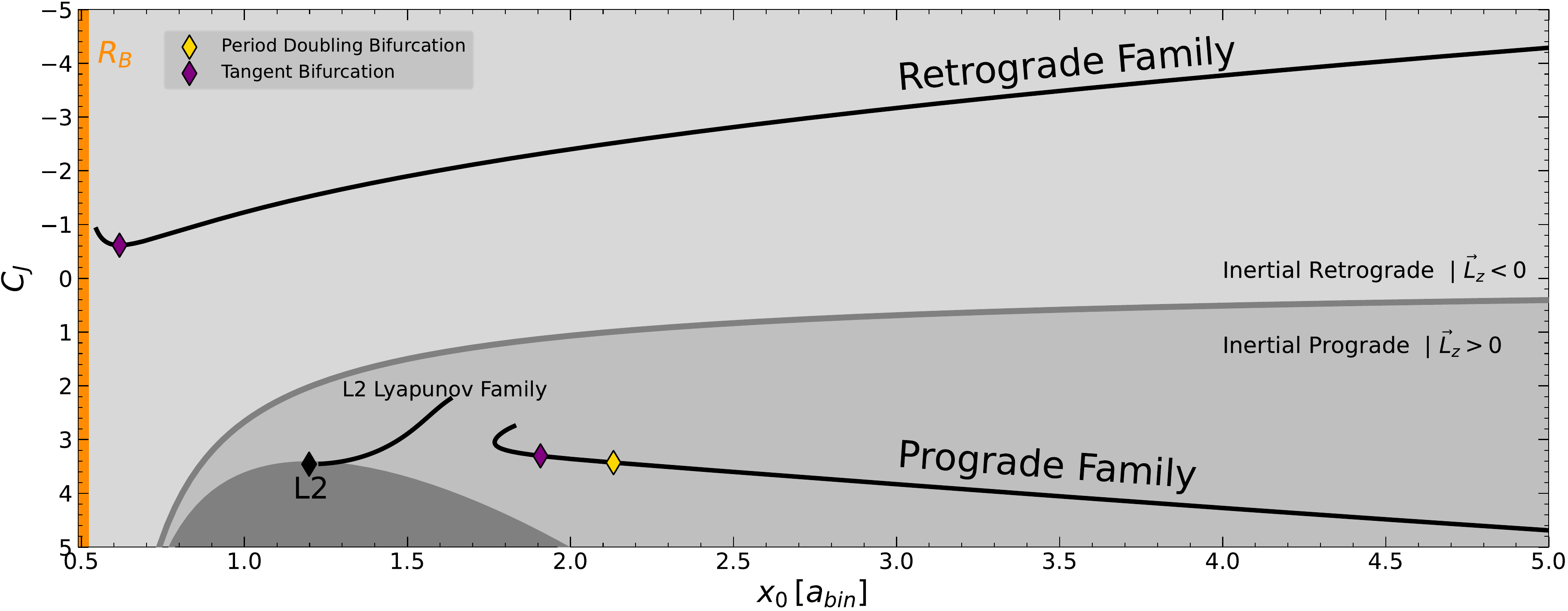}
    \caption{Families of circumbinary orbital limit cycles (black) are plotted as a function of their positive x-axis crossing and Jacobi Constant. This visualization highlights the locations of limit cycle bifurcations (diamonds) and evolution of periodic motion through state space. The negative x-axis crossing is symmetric (except for the L2 Lyapunov Family) and is not shown. The prograde and L2 planar Lyapunov families are in non-intersecting regions of the solution space.}
    \label{fig: bifplot_CP}
\end{figure*}

On the right side Figure \ref{fig: limit_pert_exp}, limit cycles are perturbed in the bifurcating eigendirection by $10^{-1}$. The post-perturbation behavior of each trajectory qualitatively agrees with the linear dynamical prediction of Floquet theory. The trajectory exterior to the bifurcation oscillates about the periodic solution, resulting in bounded motion in both the synodic and inertial frames. However, the perturbed trajectory interior to the bifurcation departs exponentially from its associated periodic solution and eventually appears ejected from the system. 

This numerical experiment of in-plane stability contextualizes our Floquet theory calculations within the scope of long-integration methods \citep{Holman1999, Doolin2011, Quarles2018, Chen2019}. We have demonstrated that trajectories in the vicinity of limit cycles with oscillatory dynamics remain bounded for several binary revolutions\footnote{See \S \ref{Sec: NbodySims} for analysis with $\sim 10^5 \, P_\textrm{bin}$ simulations.}, whereas trajectories surrounding limit cycles with exponential dynamics are ejected from the system in the same time. Hence, the bifurcating limit cycles offer a partition between the solution spaces of unbounded and bounded circumbinary trajectories. Bifurcations within the family of planar circumbinary limit cycles offers a plausible underlying dynamical mechanism to behind the widely documented instability of circumbinary trajectories close to the binary \citep{Holman1999, Doolin2011, WinnFabrycky2015, Li2016, Quarles2018, Chen2019}. 

\subsubsection{Circumbinary Solution Space for $\mu=0.5$} \label{Sec: CircSolSpace}

\begin{figure*}[t]
    \centering
    \includegraphics[width = \linewidth]{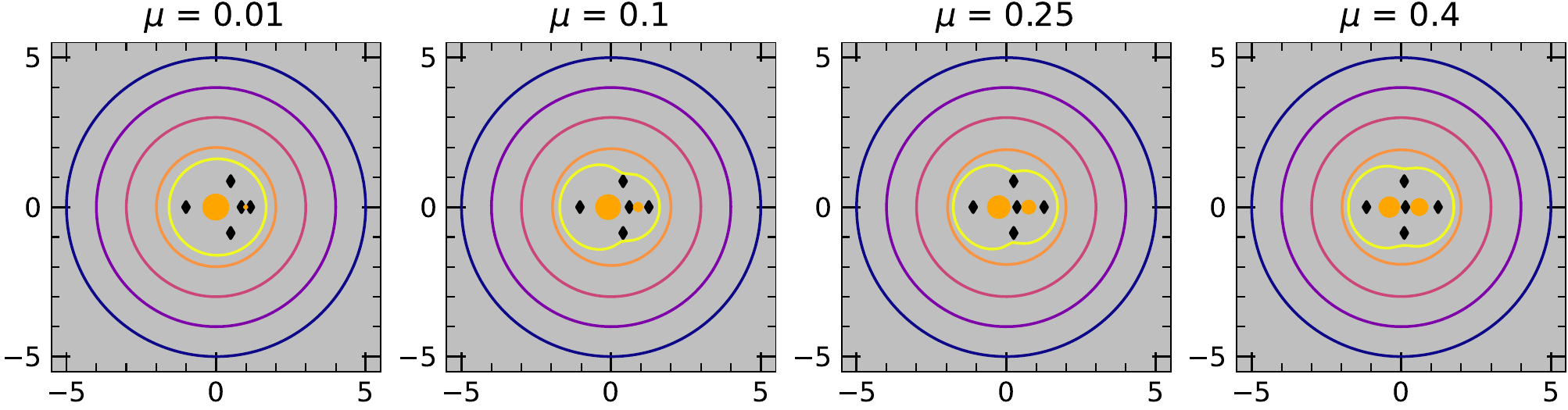}
    \includegraphics[width = \linewidth]{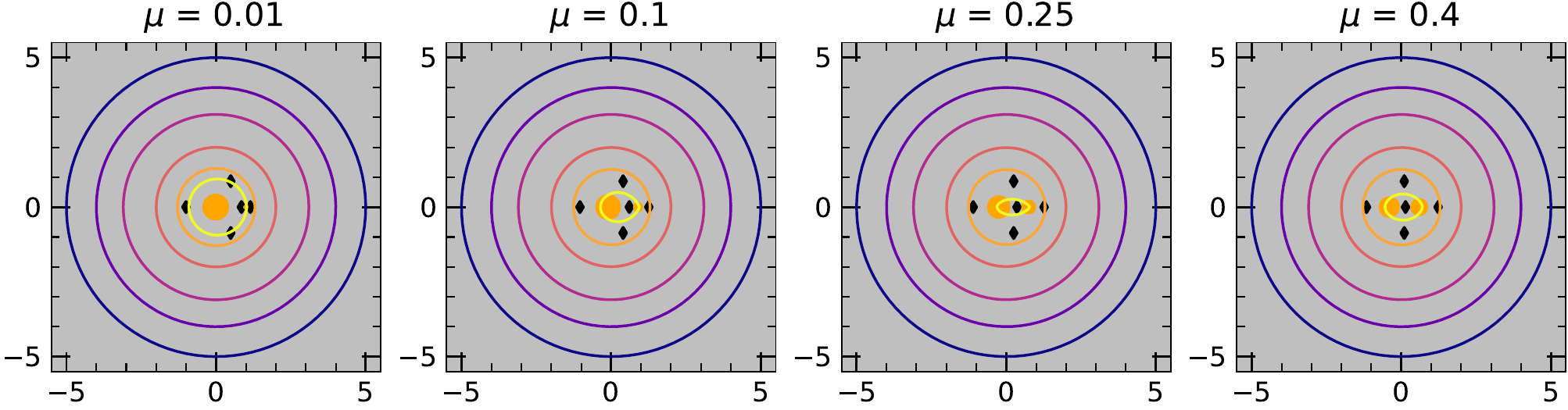}
    \caption{Sparse representations in the synodic frame of the prograde (top) and retrograde (bottom) families of periodic circumbinary orbits for selected values of the binary mass ratio $\mu$. All trajectories propagate in a clockwise direction in the synodic frame. The primaries' relative sizes are scaled with their fractional masses. The five fixed points (Lagrange points) for each system are shown. For all mass ratios, periodic trajectories beyond $2 \, a_\textrm{bin}$ retain a near-circular geometry }
    \label{fig:CB_geo_allmu}
\end{figure*}

For trajectories orbiting one primary in the CR3BP, the accessible regions of a trajectory are generally defined by its zero-velocity curves. However, for trajectories in circumbinary motion, there is no exterior and sometimes no interior boundaries to the regions of configuration space accessible to the trajectory. In short, the CR3BP provides minimal criteria for bounded trajectories in circumbinary motion. However, families of limit cycles offer an emergent dynamical structure that characterizes the typical geometric and stability properties of bounded circumbinary motion. 

To visualize the circumbinary solution space, Figure \ref{fig: bifplot_CP} presents families of periodic solutions exterior to the second mass, plotting the $+x$-axis crossing vs Jacobi Constant. The vertical axis in Figure \ref{fig: bifplot_CP} is inverted as to associate the Jacobi Constant with the more intuitive invariant framework of energy. A cross section of the equal-mass zero-velocity surface represents regions of solution space not accessible to a trajectory with real-valued velocity. The boundary between a trajectory orbiting in the prograde versus retrograde sense with respect to the inertial frame is also illustrated and agrees with our previous claims about families' direction of motion.\footnote{The boundary between prograde and retrograde is generated by solving for Jacobi Constant at a given $x$-axis crossing that results in a velocity with zero angular momentum in $\hat{z}$ with respect to the barycenter.} The synodic location of $m_2$ is denoted by the orange vertical line at $x = + 0.50$.

Figure \ref{fig: bifplot_CP}, which we refer to as a \textit{bifurcation plot}, provides a succinct way of contextualizing the the four-dimensional planar circumbinary solution space. The continuous families of periodic orbits are shown as thick black lines: the prograde family, retrograde family, and the L2 Lyapunov family.\footnote{The planar L2 Lyapunov family is an set of unstable limit cycles that orbit the second Lagrange point and provide a relevant dynamical structure for contextualizing circumbinary prograde motion.} Note that the families do not overlap or intersect in regions of state space computed. The retrograde family extends almost all the way to the binary, whereas the prograde family turns around at $x_0\approx 1.767 \, a_\textrm{bin}$. The prograde family might be limited close to the binary by the existence of the L2 fixed point and associated L2 Lyapunov family, although a detailed investigation of the interaction of these families is outside the scope of this work. The prograde family has two bifurcations --- a period-doubling bifurcation at $x_0\approx 2.131 \, a_\textrm{bin}$ and a tangent bifurcation at $x_0\approx 1.907 \, a_\textrm{bin}$ --- both of which are exterior to the inflection point where the prograde family turns around.\footnote{Solving the tangent limit cycles at the bifurcating trajectories would fill in regions of the solution space with more exotic periodic families.} The retrograde family is unimpaired as it approaches the binary due to the lack of dynamical features in its surrounding state space. The in-plane tangent bifurcation occurs at the retrograde family's minimum Jacobi Constant. Past this point, trajectories possess unbounded behavior in the plane and decreasing Jacobi Constant. The locations of circumbinary limit cycles and their bifurcations in state space provides observational constraints on the location of circumbinary motion similar to zero-velocity curves in circumstellar motion.

\section{Dynamical Behavior Across Mass Ratio} \label{Sec: MassRatios}

The Copenhagen Problem (\S\ref{Sec: CP}) serves as limiting case of the CR3BP. In this section, we extend the dynamical systems approach to investigate the emergence of dynamical features for a variety of binary system mass ratios. The mass ratios presented in this section, $\mu \in [0.01,0.50]$, cover the observed mass ratios of the NASA Kepler and TESS CBP host binaries \citep{WinnFabrycky2015, Kostov2021}. These mass ratios are distinct from $\mu < 0.01$, where the CR3BP periodic families change structure rapidly and converge on the Hill Problem \citep{BosanacHowellFischbach2015}.

\subsection{Orbital Geometries} \label{Sec: allmu_orbgeo}

We computed the CR3BP circumbinary prograde and retrograde families using the single shooting algorithm and pseudo-arclength continuation techniques presented in Sections \ref{Sec: DiffCorr} and \ref{Sec: InitContFam}. A sparse representation of the families for select mass ratio values is presented in Figure \ref{fig:CB_geo_allmu} for a selection of $x$-axis crossings (blue, purple, magenta, rose, and orange lines) and also for the innermost calculated orbits (yellow) in each family.

The periodic trajectories far from the barycenter are nearly identical for all mass ratios. For trajectories with $x_0 \lesssim 2 \, a_\mathrm{bin}$, slight differences in geometry emerge between mass ratios. In the prograde families, the location of the cusp for the innermost trajectory follows the L4 and L5 fixed points. For mass ratios $\mu \in [0.2, 0.5]$ we found a turning point within the circumbinary prograde trajectories, after which continuing the family results in high-energy, rapidly precessing orbits. In the retrograde family, distinct visual differences between the periodic trajectories across mass ratio only appear interior to the Lagrange points. The innermost retrograde periodic trajectories all flatten along the $x$-axis, except for $\mu = 0.01$, which retains a near-circular geometry. In general, the periodic trajectories across mass ratios possess subtle changes to the orbital geometry, with trajectories with $x_0 > 2 \, a_\mathrm{bin}$ minimally changing as a function of $\mu$.

\begin{figure}[t]
    \centering
    \includegraphics[width = \linewidth]{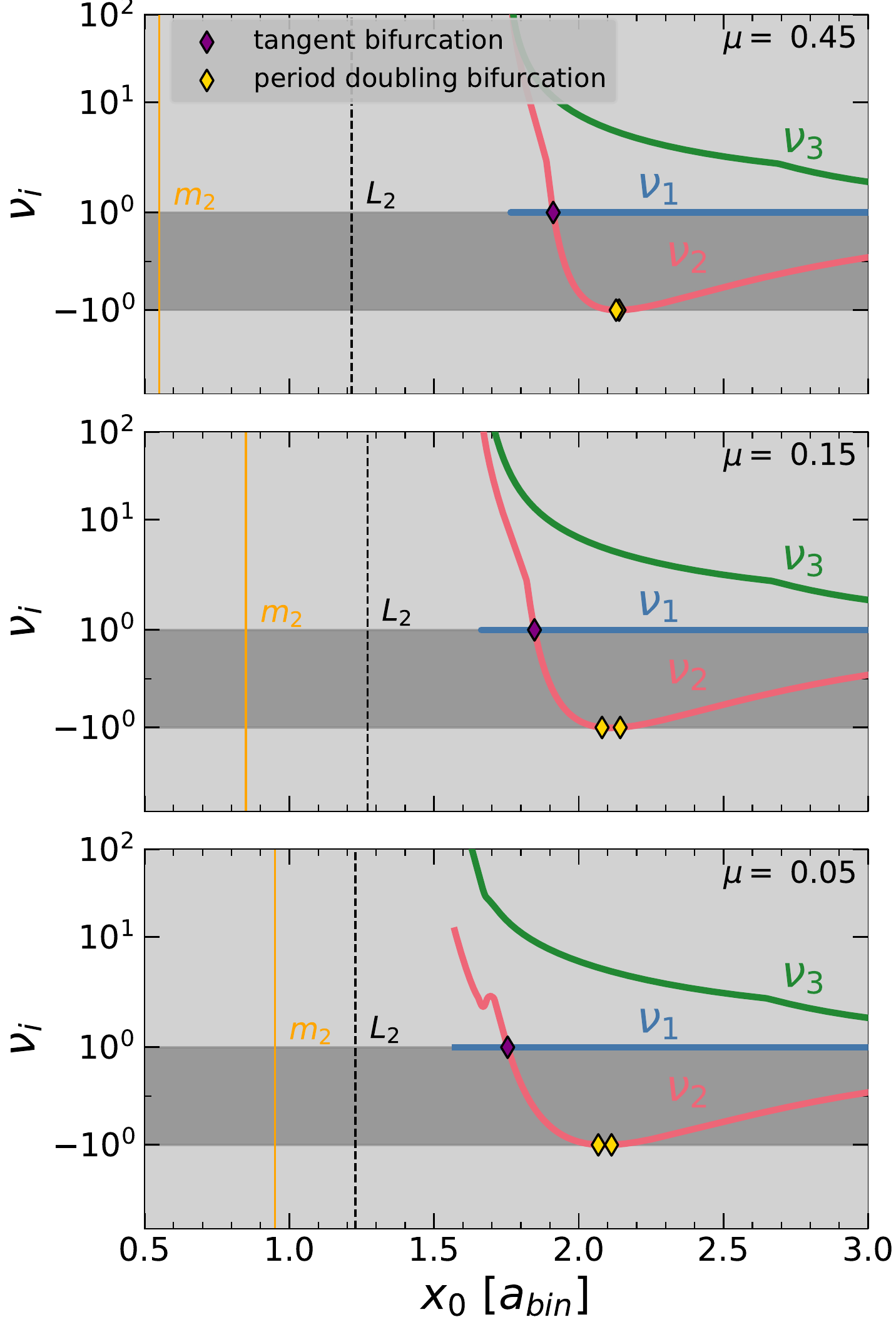}
    \caption{Same as Figure \ref{fig: mu05_nuvx} left panel, but for a variety of stellar mass ratios; $\mu = [0.05, 0.15, 0.45]$.  The positions of the secondary star ($x_0 = 1-\mu$, orange line) and L2 (black dashed line) are indicated. Note that for $\mu < 0.50$, there are two period-doubling bifurcations on either side of a region where in-plane perturbations lead to exponential growth. The locations of the period-doubling and tangent bifurcations change as a function of stellar mass ratio. }
    \label{fig:munu_pro}
\end{figure}
\begin{figure}[t]
    \centering
    \includegraphics[width = \linewidth]{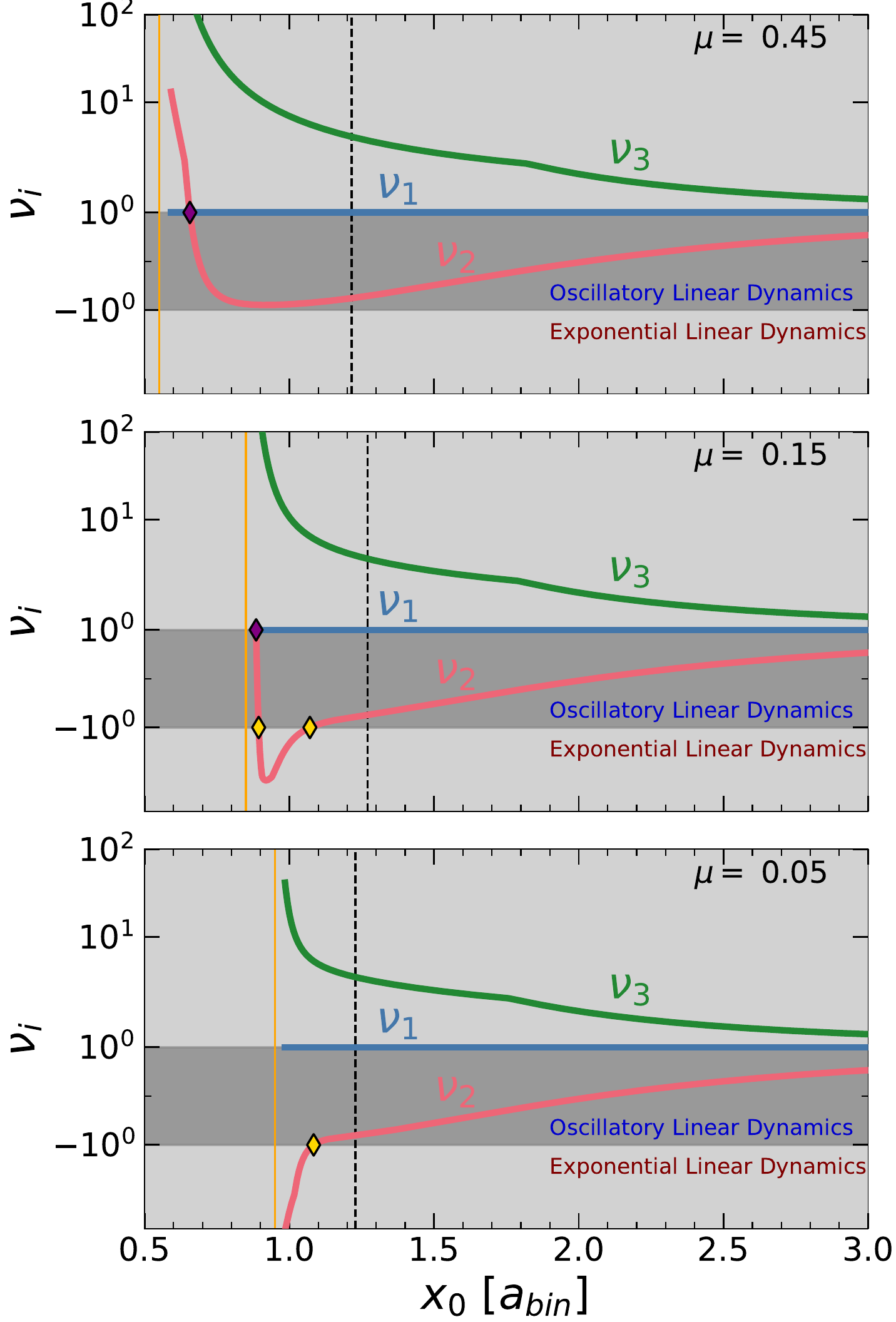}
    \caption{Same as Figure \ref{fig:munu_pro}, but for retrograde orbits, analogous to the right panel of Figure \ref{fig: mu05_nuvx}.  We identified a tangent bifurcation corresponding to an innermost stable orbit for $\mu=0.45$ and $\mu=0.15$, but not for $\mu=0.05$.  Note the emergence of a pair of period-doubling bifurcations between $\mu=0.45$ and $\mu=0.15$.}
    \label{fig:munu_ret}
\end{figure}
\subsection{Dynamical Evolution Across Mass Ratio}

In this section, we extend the dynamical systems approach of \S\ref{Sec: BifDet} to the computed retrograde and prograde periodic families presented in \S\ref{Sec: allmu_orbgeo} to better understand how the mass ratio affects circumbinary orbital dynamics.

\subsubsection{Bifurcation trends in mass ratio} \label{Sec: bif_munu}

For a given mass ratio system, we identified bifurcations in the stability of circumbinary motion by computing the parameter $\nu_i$ (see Equation \ref{Eq: Nu_def}) across the families of circumbinary limit cycles. Bifurcations occur within a family at the point $\nu_i = \pm 1$ instantaneously. Figures \ref{fig:munu_pro} and \ref{fig:munu_ret} illustrate how the bifurcation parameters $\nu_i$ evolve with the binary mass ratio $\mu$. As in \S\ref{Sec: BifDet}, $\nu_1$ and $\nu_2$ correspond to eigendirections in the plane of the binary ($x, y, \dot{x}, \dot{y}$) and $\nu_3$ out-of-plane ($z, \dot{z}$). 

In the prograde families, there are two successive period-doubling bifurcations as limit cycles approach the binary, Figure \ref{fig:munu_pro}. The solution space along the family between the period-doubling bifurcating limit cycles exhibits exponential linear dynamics. These dynamics are characterized by trajectories unstable to perturbations in the plane of the binary. This unstable region emerges for all computed values of $0.01 \leq \mu < 0.50$. Our identification of this region correlates to a band of planar instability illustrated in \citep{BosanacHowellFischbach2015}'s Figure 5a for limit cycles with synodic periods $T = \approx 9.42$ in mass ratios $10^{-5} \leq \mu < 0.50$. Interior to the period-doubling bifurcations, we detect tangent bifurcations in $\nu_2$. Past the tangent bifurcating limit cycles, prograde trajectories are generally unbounded in the plane. However, \citet{BosanacHowellFischbach2015} (Fig 5a) finds a narrow band of in-plane stable trajectories that possess looping geometries for $\mu \in [0.1, 0.5]$. Overall, the bifurcation features across mass ratios $\mu \in [0.05, 0.50]$ generally resemble the Copenhagen problem, with the addition small regions of instability between period-doubling bifurcations. This characterization of the prograde circumbinary CR3BP solution space is also shown by \citet{BosanacHowellFischbach2015}'s Figure 5 `exclusion plot' visualizing qualitative stability as a function of synodic period and system mass ratio. 

The retrograde family bifurcations differ across mass ratios more significantly than their prograde counterparts, as detailed in Figure \ref{fig:munu_ret}. At high mass ratios, $\mu \geq 0.33$, we measure a single tangent bifurcation occurring close to the location of the secondary mass (Fig. \ref{fig:munu_ret} top panel). Recall that trajectories interior to the tangent bifurcation are unbounded in the plane of the binaries (\S\ref{Sec: BifDet}). Between mass ratios $0.13 \ge \mu \ge 0.32$, a pair of period-doubling bifurcations emerges, limiting the regions of bounded motion exterior to the tangent bifurcation. Around $\mu = 0.13$, $\nu_2$ approaches an infinite slope (Fig. \ref{fig:munu_ret} middle panel). For $\mu < 0.13$, we only detect one period-doubling bifurcation (Fig. \ref{fig:munu_ret} bottom panel). In this mass regime, $\nu_2$ no longer has the inflection that produces another period-doubling and tangent bifurcations. Therefore, circumbinary systems with this primary mass ratio fundamentally have a different dynamical structure from the Copenhagen problem.

\subsubsection{CR3BP Solution Space for Various Mass Ratios}

\begin{figure*}[t]
    \centering
    \includegraphics[width = \linewidth]{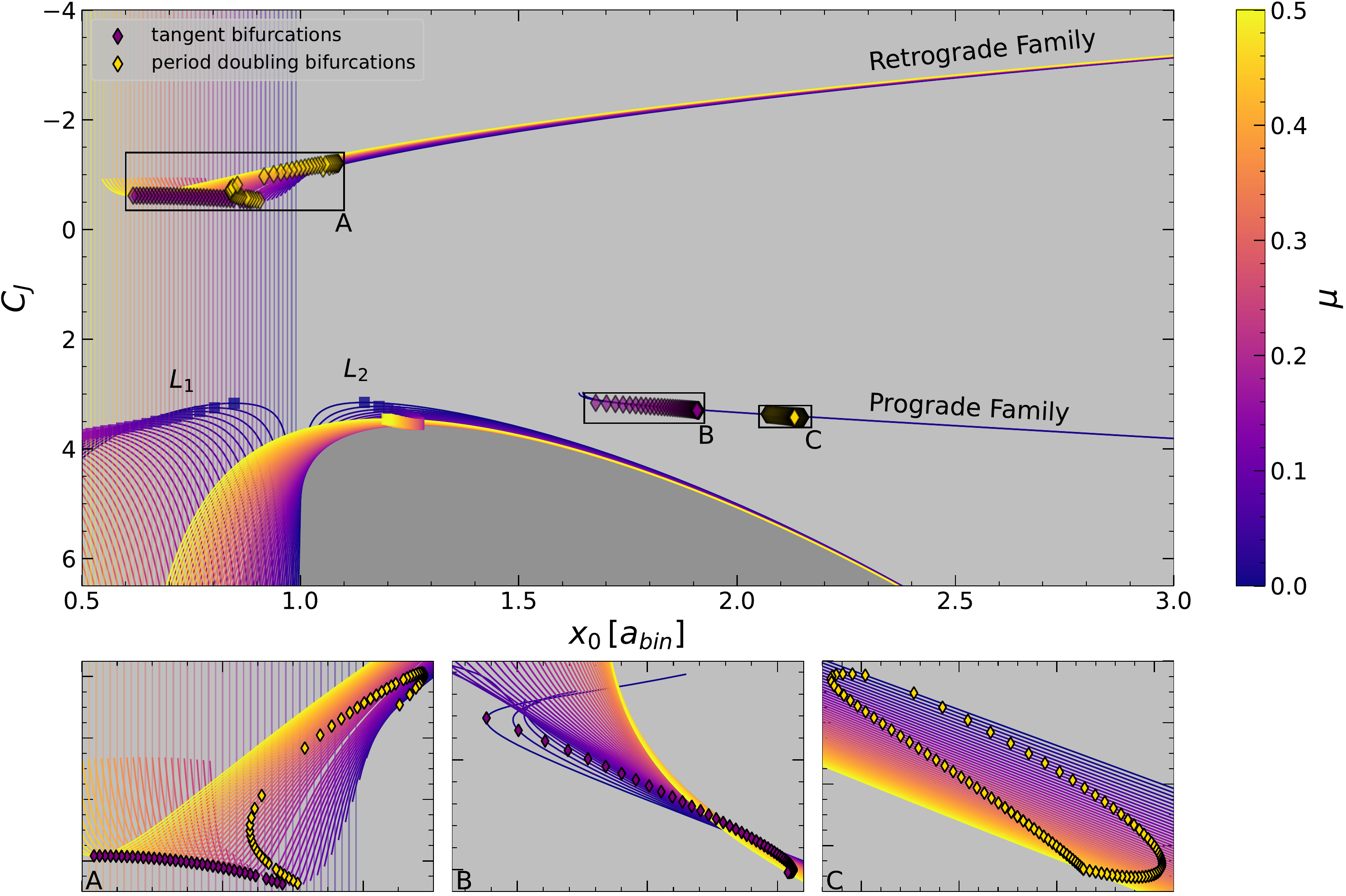}
    \caption{Top: same as Figure \ref{fig: bifplot_CP}, but with different axis limits, and for a variety of stellar mass ratios ($\mu$, colors). The retrograde and prograde orbital families are shown (thick lines), as are cross sections of the zero-velocity contours (thin curved lines) and locations of the secondary star (thin vertical lines).  Bifurcations (diamonds) are shown in each family. The L2 Lyapunov families are excluded for clarity, although the L1 and L2 Lagrange points are indicated.  Bottom: regions of bifurcations in the retrograde families (A), the tangent bifurcation in the prograde families (B), and the period-doubling bifurcations in the prograde families (C).}
    \label{fig:cr3bp_bifur_allmu}
\end{figure*}

The computation of retrograde and prograde circumbinary periodic families for mass ratios $\mu \in [0.01, 0.50]$ generates a large volume of data. The solution space of six-dimensional limit cycles across the mass parameter $\mu$ is shown in Figure \ref{fig:cr3bp_bifur_allmu}, an extension of Figure \ref{fig: bifplot_CP} for a variety of mass ratios. Figure \ref{fig:cr3bp_bifur_allmu} illustrates the similarities in the solution space for retrograde and prograde limit cycles with varying mass ratios. Each family monotonically converges towards a similar solution with increased distance from the binary. This is consistent with an understanding that the first term of the binary potential expansion does not contain the mass ratio.

The solution space of the retrograde families remains relatively intact until near the first period-doubling bifurcations located in low-mass ratio families at $x_0 \approx 0.87$. Panel A of Figure \ref{fig:cr3bp_bifur_allmu} shows the progression of bifurcations in retrograde families across mass ratio. As in Figure \ref{fig:munu_ret}, families of high mass ratios, $\mu \geq 0.34$, only have a tangent bifurcation. At $\mu \approx 0.33$, a pair of period-doubling bifurcations emerge and continue until $\mu \approx 0.13$ where the interior period-doubling and tangent bifurcations collide. Figure \ref{fig:cr3bp_bifur_allmu} also visually reveals that the tangent bifurcations are located at the maximum valued Jacobi Constant solution within the retrograde families. While not explicitly solved for in this investigation, the collision between the tangent and period-doubling bifurcations may delineate the termination of maximum Jacobi Constant solutions within the retrograde family. 

Prograde limit cycles remain similar interior to the period-doubling bifurcations (Panel C), and differentiation only occurs closer to the tangent bifurcations (Panel B). In mass ratios $\mu \in [0.01, 0.04] \cup [0.11, 0.50]$ we identify turning points along the prograde family resulting after which limit cycles move outward along $\hat{x}$. For $\mu = 0.01$, the turning point is near the tangent bifurcation whereas for higher values of $\mu$, the turning point is interior to the tangent bifurcation and thus represents a solution space unstable to linear perturbations. Panels B and C of Figure \ref{fig:cr3bp_bifur_allmu} details bifurcations along the prograde families of varying mass ratios. In Panel B, we observe a slight outward progression of the tangent bifurcations until $\mu \approx 0.42$. Between $0.2 \lessapprox \mu \leq 0.5$, the $x_0$ location of the tangent bifurcations is bounded between $[1.872, 1.193]$. The relatively small change in locations of tangent bifurcations over a large range of mass ratios is indicative of how close-in circumbinary motion is similarly effected by the presence of two massive bodies of varying mass ratio. In low mass ratio families, we observe a rather drastic change in minimum $x$-crossing locations between $\mu = 0.04 \rightarrow \text{min}(x_0) = 1.676$ and $\mu = 0.05 \rightarrow \text{min}(x_0) < 1.571$. However, the progression of tangent bifurcations is continuous across all mass ratios and monotonically decreases in Jacobi Constant with smaller mass ratios. Panel C shows the paired evolution of period-doubling bifurcations in the prograde families. We observe the pair originates in the $\mu = 0.5$ family at $x_0 = 2.1318$. The exterior bifurcation reaches a maximum $x_0 = 2.1520$ at $\mu = 0.27$. The interior period-doubling bifurcation has a minimum $x_0 = 2.0671$ at $\mu = 0.06$. The pair appears to be converging and would collide at $x_0 \approx 2.08$ as $\mu \rightarrow 0$ -- of course, no bifurcations should exist at $\mu = 0$. The widest range of $x_0$ values encompassed between the period-doubling bifurcations occurs at $\mu = 0.13$, $\Delta x_0 = 0.0634$. Our computations and bifurcation analysis of the prograde family are in agreement with \citet{BosanacHowellFischbach2015}'s computation of circumbinary planar stability in the same mass ratio domain.

\section{Critically Stable Trajectories} \label{Sec: a_cr}

Unlike their single-star counterparts, circumbinary planetary systems have an innermost stable trajectory that strictly limits bounded motion. Material that exists interior to this critically stable orbit is quickly ejected from the system. The distance from the barycenter of the innermost stable trajectory, $\acr$, is considered a fundamental property of a binary's planetary architecture \citep{WinnFabrycky2015}. For this reason, $\acr$ is a frequent computation attempted across a range of systems and dynamical models \citep{Holman1999, Doolin2011, Quarles2018, Chen2020}. A binary's eccentricity and mass ratio are the dominant dynamical parameters that influence the location of the system's innermost stable trajectory. The bifurcation plot (Figure \ref{fig:cr3bp_bifur_allmu}) provides a broad illustration of the circumbinary orbital families at a given mass ratio for circular binary. However, for comparing results between mass ratios and other investigations, we are most interested in the limit cycles within the families that define the boundaries between otherwise dynamically similar solutions. For the application of our results to a broader context, we define two types of limit cycles with the following characterizations:
\begin{enumerate}
    \item[] \textbf{critically stable trajectory} -- a bifurcating limit cycle that defines the boundary between solution spaces of qualitatively different dynamics 
    \item[] \textbf{innermost stable trajectory} -- the critically stable limit cycle that is closest to the binary (denoted $\acr$ in this work)
\end{enumerate}

In this investigation we have exactly determined the innermost stable trajectory for prograde and retrograde orbits permitted by the CR3BP across the mass ratios $\mu \in [0.01, 0.50]$ through bifurcation analysis. We also have identified a number of critically stable trajectories that are not the innermost stable trajectory but still define the boundaries between regions of stable and unstable planar dynamics. Our generalized results permit a direct comparison of the dynamical systems approach to methods of long-term integration for measuring $\acr$ in binary systems. 

\subsection{Prior approaches for identifying the innermost stable trajectory}
Before comparing the results for an innermost stable trajectory, it is worth reviewing the models and approaches applied to computing $\acr$. \citet{Holman1999} applied the Elliptical Restricted Three-body Problem (ER3BP) to model the long-term behavior of test particles on initially prograde, circular Keplerian orbits. They determined $\acHW$ by the initial distance required for eight equally spaced particle mean anomalies to survive for $10^4 \ P_\textrm{bin}$. \citet{Doolin2011} also applied the ER3BP to initially circular Keplerian particles for $10^4 \ P_\textrm{bin}$, building on \citet{Holman1999} by (1) doubling the sampling resolution of initial test particle locations, and (2) examining particles in retrograde and non-planar trajectories. \citet{Doolin2011} plotted the survivability of initial conditions in their simulation, but did not explicitly list the values or open-source their results. To determine $\acDB$ we visually examined Figure 14 of \citet{Doolin2011} and estimated the values and errors of the distance of their innermost stable orbits. \citet{Quarles2018} employed an N-body simulation using a sympathetic integrator from the \texttt{mercury} \citep{Chambers2002} package to test the survivability of Jupiter-mass bodies on prograde planar trajectories with initial Keplerian circular velocities for $10^5$ P\textsubscript{bin}. \citet{Quarles2018} used ten times the sampling resolution of \citet{Holman1999} in initial particle locations, mean anomalies, and binary mass ratio values. \citet{Quarles2018} applied the same criteria as \citet{Holman1999} for determining $\acQ$ -- the initial semi-major axis for which particles at all initial mean anomalies survived the length of integration. \citet{Chen2020} also investigated the stability of planar and polar circumbinary orbits with initially Keplerian circular velocities through long N-body integration via a second-order symplectic Wisdom Holman integrator in \texttt{rebound} \citep{Rein2012}. \citet{Chen2020} did not test $e_{bin} = 0$ or provide explicit results for $\acr$.

\subsection{Comparing critically stable trajectories between investigations}

\begin{figure*}[t]
    \centering
    \includegraphics[width = \textwidth]{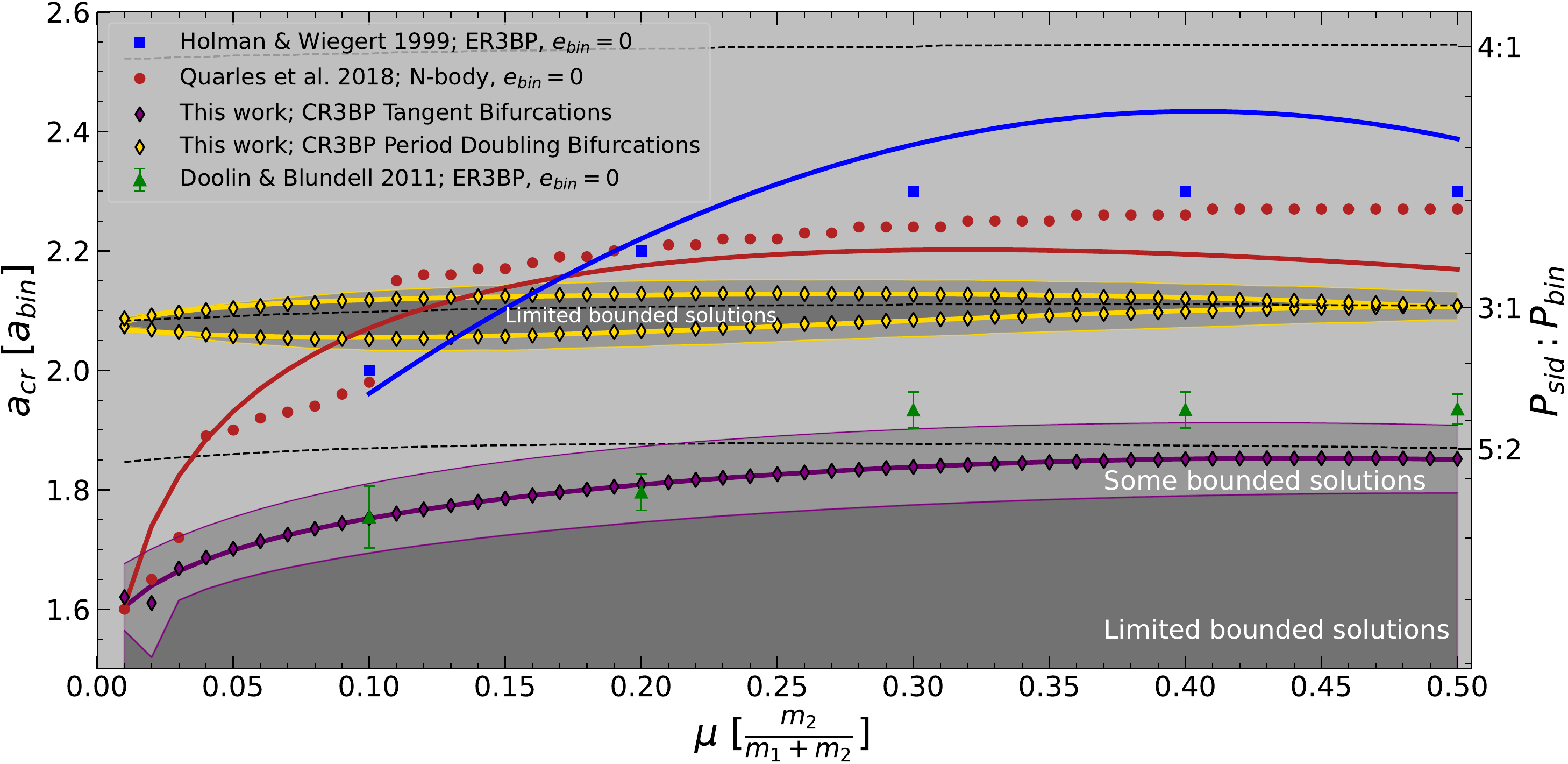}
    \caption{Critical distances from the barycenter ($\acr$) versus mass ratio ($\mu \in [0.01,0.50]$) for bounded planar circumbinary trajectories in the CR3BP prograde direction. Notable sidereal resonances (dashed line) are shown as a function of binary mass ratio. Determinations of $\acr$ from previous investigations are plotted (various shapes) along with their fitted models when available (solid lines, see \S\ref{Sec: parfit}). The geometric semi-major axis of the critical bifurcating trajectories computed in this work (diamonds) are fit with a parametric model (Equation \ref{Eq: acr_fit}, thick lines), with purple representing tangent bifurcations and yellow representing period-doubling bifurcations. The apoapsis and periapsis of the critical stable orbits are shown (thin lines). Our stability analysis of the critical trajectories yields regions where no stable orbits exist (double-hashed shaded regions).}
    \label{fig:acr_pro}
\end{figure*}

\begin{figure*}[t]
    \centering
    \includegraphics[width = \textwidth]{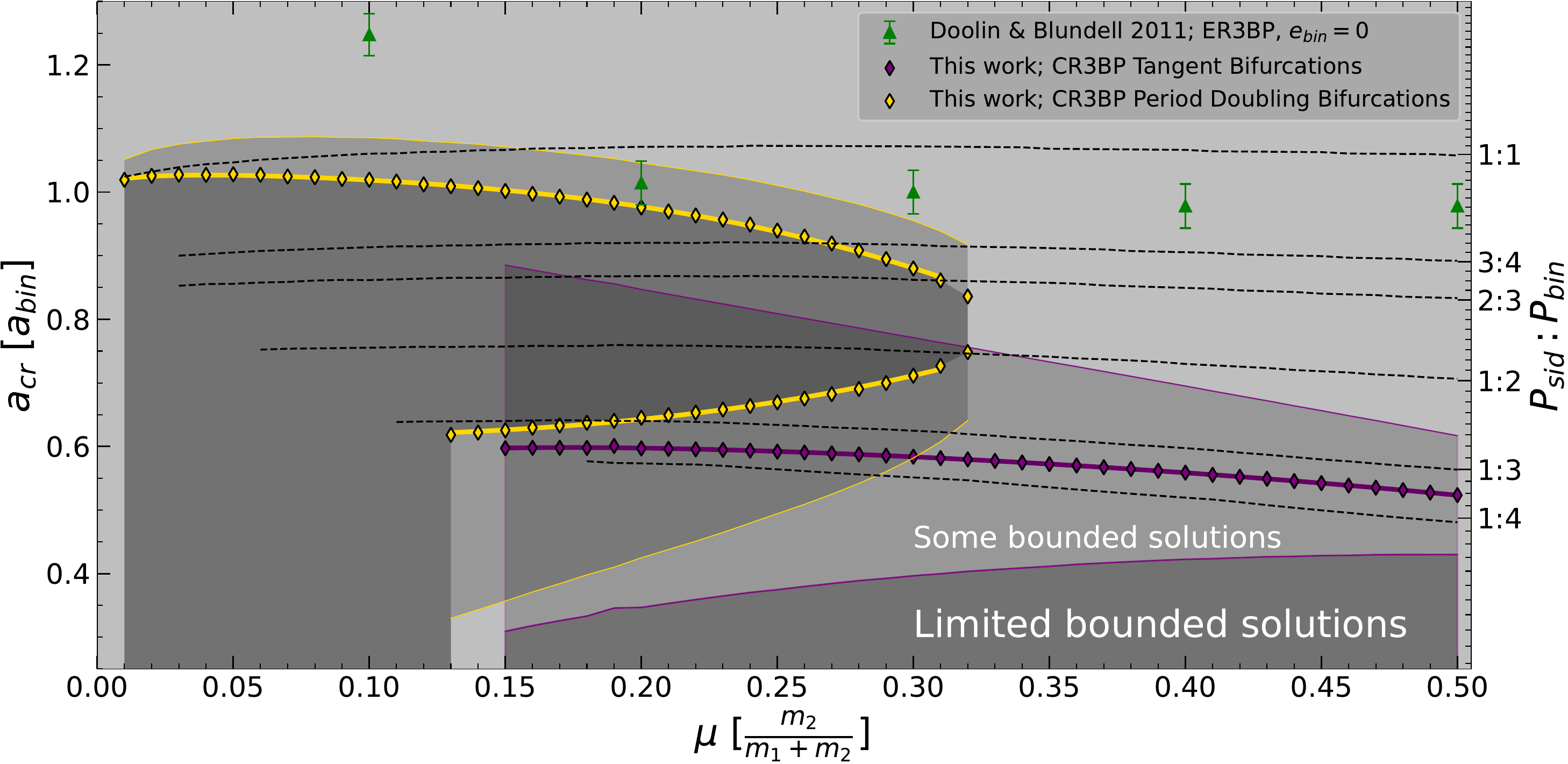}
    \caption{Same as Figure \ref{fig:acr_pro}, but for retrograde trajectories. Note, some stable circumbinary retrograde trajectories have shorter orbital periods than their binary as a consequence of their orbital geometries $(a_\textrm{geo}<a_\textrm{bin})$. See Figures \ref{fig:ret_fam} and \ref{fig:CB_geo_allmu}.}
    \label{fig:acr_ret}
\end{figure*}

We now consider prior long-integration and our dynamical systems theory results for the location of the innermost stable trajectory, $\acr$, of a circumbinary planet around a circular binary for mass ratios $\mu \in [0.01, 0.50]$. Our dynamical systems results provide critically stable trajectories that separate regions of unbounded and bounded dynamics in the six-dimensional state space. To generalize these results for comparison against previous works, we adopt a critically stable trajectory's geometric semi-major axis, $a_\textrm{geo} = (r_a + r_p)/2$, as our measurement of the representative distance from the barycenter. 

Figure \ref{fig:acr_pro} compares results for $\acr$ and critically stable trajectories in prograde planar trajectories across four investigations. Fitted expressions for $\acr$ are plotted over the domain of their original data when available.\footnote{Methods of $\acr$ parameterization are compared in \S\ref{Sec: parfit}.} Our analysis yielded values of $\acr$ that are equal to or interior to values found by previous studies for all values of $\mu$. We also find regions of unbounded planar dynamics exterior to the innermost stable trajectory established by a pair of critically stable trajectories. These critically stable trajectories are a consequence of the pair of period-doubling bifurcations within the prograde families -- thus we refer to these unstable regions as the \textit{period-doubling exclusion zone}. A four-coefficient model, Equation \ref{Eq: acr_fit}, approximates how the semi-major axes of critically stable trajectories vary with stellar mass ratio (Table \ref{Tab: acr_coeff}).

Each investigation found $\acr$ to be nearly constant over the domain $\mu \in [0.30, 0.50]$. Interestingly, while \citet{Holman1999} (blue squares) and \citet{Quarles2018} (red dots) implemented different models (restricted and N-body, respectively) and sampling resolutions, their results for $\acr$ generally agree. Notably, \citet{Quarles2018} innermost stable trajectories straddle (at $\mu\approx 0.11$), but do not fall within, the period-doubling exclusion zone we computed (yellow diamonds). Meanwhile, \citet{Doolin2011}'s long-term integrations (green triangles) yielded innermost surviving trajectories that are consistent with the innermost bounded trajectories we computed (purple diamonds). Note that, in the limit, $e_\mathrm{bin}=0$, the results of \citet{Holman1999}, \citet{Doolin2011}, and our work are from identical dynamical models (restricted circular), whereas the results of \citet{Quarles2018} treat the planet mass as non-zero.

Figure \ref{fig:acr_ret} shows the planar, retrograde innermost stable trajectories found by \citet{Doolin2011} and our retrograde critically stable trajectories. As a consequence of the changing bifurcation landscape in the retrograde families, our dynamical systems results produce a nuanced picture of the innermost stable trajectory. At high mass ratios, the retrograde exclusion zone is non-existent, and innermost stable trajectory is set by the tangent bifurcating periodic trajectory. Our disagreement from $\acDB$ at large mass ratios is likely explained in one of two ways. First, \citet{Doolin2011} also did not explicitly state the minimum sampled distance, so it is possible they did not initialize retrograde orbits as close to the binary as we did in our approach. Second, if \citet{Doolin2011} did sample close to the binary, perhaps their 2BP initial conditions were insufficient for generating bounded motion in the three-body dynamical environment, whereas our approach of differentially correcting trajectories to converge on a closed orbit guarantees initial conditions that produce bounded motion. Between $0.13 \leq \mu \leq 0.32$ the solution space directly exterior to the innermost stable trajectory becomes filled in by a period-doubling exclusion zone. This exclusion zone rapidly decreases the permissible solution space for bounded motion near the tangent bifurcation as mass ratio decreases. For $\mu \lesssim 0.12$, there is only one period-doubling bifurcation (recall Figure \ref{fig:munu_ret}), and all orbits interior to this bifurcation are unstable to perturbations. For these low mass ratios, the innermost stable trajectories found by \citet{Doolin2011} are consistent with the values of $\acr$ we determined from our bifurcation analysis. We measure $\acLW$ as fairly constant in retrograde trajectories for low mass ratios ($\mu \leq 0.12$) and expect $\acLW \rightarrow 1$ as $\mu \rightarrow 0$ -- consistent with the 2BP.

Mean motion orbital resonances between the binary and CBP are often treated as stabilizing or destabilizing dynamical features for the planet's trajectory. Along with critical semi-major axes, Figures \ref{fig:acr_pro} and \ref{fig:acr_ret} plot the equivalent semi-major axes of notable sidereal resonances as a function of the binary mass ratio. These sidereal resonances are computed using,
\begin{equation}
    \frac{1}{P_\textrm{sid}} = \left |\frac{1}{P_\textrm{bin}} - \frac{1}{P_\textrm{syn}} \right |
\end{equation}
where $P_\textrm{bin}$ is the period of the binary, $P_\textrm{syn}$\footnote{Previously referred to as $T$ in \S \ref{sec:dynamical_background}.} is the limit cycle's synodic period, and $P_\textrm{sid}/P_\textrm{bin} = m/n$ where $m, n \in \mathbb{Z}$. Compared to two-body orbits at equivalent semi-major axes, the circumbinary trajectories have shorter periods -- consequently, circumbinary resonances are farther from the barycenter than their two-body counterparts. For instance, Figure \ref{fig:acr_pro} shows the $3:1$ circumbinary resonance lies between the period-doubling bifurcations. If we were to assume the stellar mass is concentrated at the barycenter and apply Kepler's Third Law ($P \propto a^{3/2}$), the $3:1$ resonance would appear to be at $2.080 \, a_\textrm{bin}$, which is interior to the exclusion zone and inconsistent with our dynamical computation of resonance.

Counter-intuitively, the critically stable trajectories do not always seem to correspond to low-order mean motion resonances.  For instance, the prograde innermost stable trajectory does not have a clear connection to the $5:2$ resonance over all mass ratios. Also, Figure \ref{fig:acr_ret} does not clearly show a connection between low mean motion resonances and the retrograde critically stable trajectories. From these results, it is difficult to establish a straightforward relationship between Floquet stability and mean-motion resonance. Further analysis of N-body simulations sampled from periodic trajectories may enhance our understanding of this relationship.

\begin{figure*}[t]
    \centering
    \includegraphics[width = \linewidth]{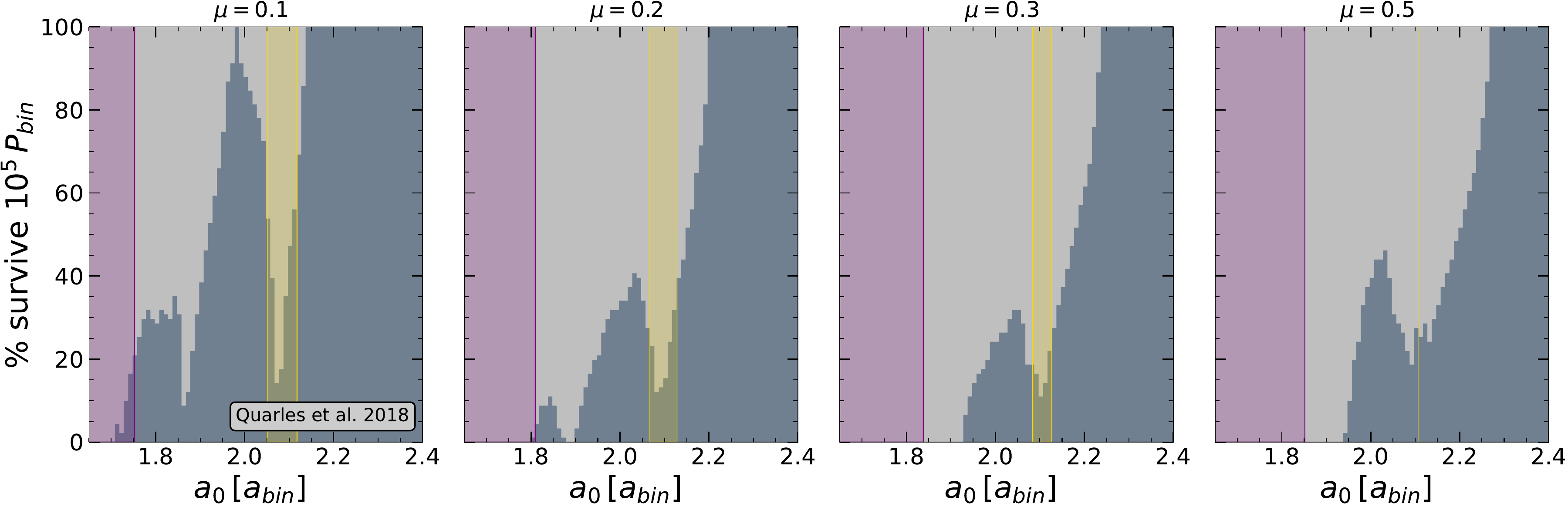}
    \includegraphics[width = \linewidth]{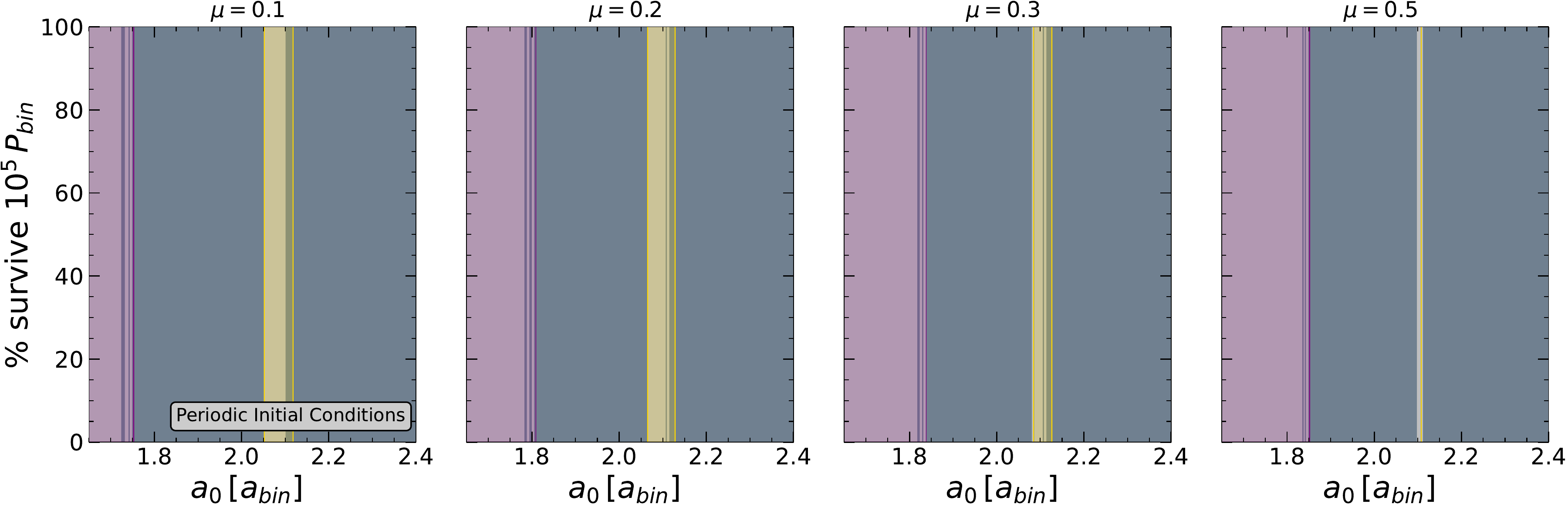}
    \caption{Top panel: Simulation results from \citet{Quarles2018}\footnote{http://doi.org/10.5281/zenodo.1174228} measuring the survivability of massive planets ($m_p \approx 1 M_J$) on initially Keplerian circular orbits for $10^5 \ P_\textrm{bin}$ around an initially zero eccentricity binary. The calculated geometric semi-major axis of critically stable trajectories from Floquet theory overlay the trajectories' semi-major axis. 
    Bottom panel: Replication of top panel but using CR3BP periodic initial conditions. In both panels, the period-doubling exclusion zone (gold) falls between a significant gap in survived trajectories. Very few simulated trajectories survive interior to the tangent bifurcating trajectory (purple). All periodic initial conditions survive in the predicted stable region, whereas only an enhancement of circular Keplerian initial conditions survive in the same region. }
    \label{fig: acr_hist}
\end{figure*}

\subsection{Comparison to long N-body results} \label{Sec: NbodySims}

To interpret the apparent discrepancy between the prograde innermost stable trajectories computed by \citet{Quarles2018} and this work, we compared the results of their open-sourced individual N-body trials to our computed regions of unbounded dynamics, Figure \ref{fig: acr_hist}. At all mass ratios of a circular binary, the \citet{Quarles2018} simulations produced a deficit in surviving trajectories that coincides with the semi-major axes of our period-doubling exclusion zone. Interior to this gap, there is an increase in survived trajectories before the innermost distance at which all are ejected. Therefore, the reason for discrepancy in $\acr$ partially stems from \citet{Quarles2018} defining the innermost stable trajectory as the smallest initial radius for which 100\% of their N-body trials survive. Our analysis predicts stable trajectories exist interior to $\acQ$, and indeed, \citet{Quarles2018} produced long-lived orbits interior to their published value of $\acr$. While there is an increase in survived trajectories interior to the exclusion zone, only simulations at low mass ratios achieve a $100\%$ survival rate for trajectories between the tangent bifurcation and interior exclusion zone boundary. Close to the binary, the periodic initial conditions that guarantee bounded motion deviate substantially from a circular Keplerian solution in high mass ratio systems (see \S\ref{Sec: CPPerGeo})). As a consequence, \citet{Quarles2018} application of circular Keplerian velocities may have not been sufficiently close to the solution space of bounded dynamics that exists between the tangent bifurcation and exclusion zone. 

To test the accuracy of stability predictions from Floquet theory, we simulated $m_p = 0.2 \, M_J$ planets initialized on CR3BP periodic trajectories for $10^5 \, P_\textrm{bin}$. While we do not expect physical CBPs to reside on precise periodic trajectories, this choice of initialization is most comparable with past investigations applying strictly circular initial conditions for survivability tests \citep{Holman1999, Doolin2011, Quarles2018}. We propogated $1,000$ trajectories with initial positions $x_0 \in [1.6, 2.5]$ for mass ratios $\mu \in [0.1, 0.2, 0.3, 0.5]$ using the \texttt{rebound} IAS15 integrator \citep{Rein2012}. Trajectories that reached a distance of $1.1$ times their initial separation from the barycenter were considered to be unstable and thus unlikely to survive in the system on observable time scales. The second row in Figure \ref{fig: acr_hist} displays the percentage of trajectories that survived by their geometric semi-major axis. A trajectory's geometric semi-major axis was computed by sampling the orbit throughout the integration. Our Floquet theory computations for regions of circumbinary stability in the CR3BP agree with the results of long-term numerical integrations. Notably, 100\% of trajectories survive in the stable region between the exclusion zone and the innermost stable trajectory, contextualizing the enhancement of surviving trajectories in the same region produced by \citet{Quarles2018}. The long-term N-body behavior of $0.2\, M_J$ planets with orbits initialized on the CR3BP limit cycles is consistent with our predictions from Floquet theory (\S\ref{sec:dynamical_background}). However, characterizing the linear theory's applicability to a broader array of non-idealized planetary trajectories will be critical for drawing comparisons to observed systems. 

\begin{deluxetable*}{llccl}[t]
\tablecaption{Comparison of functional forms and fitness for equations to critical stable radii in a circular binary.\label{tab: acr_fits}}
\tablehead{\colhead{Paper} & \colhead{Fit Equation} & \colhead{Parameters} & \colhead{Mass Ratios} &\colhead{Relative Error, $\bar{\sigma}$} } 
\startdata
\cite{Holman1999}  & $c_1 + c_2\mu + c_3\mu^2$                     & $3$ & $5$  & $6\times10^{-2}$ \\
\cite{Quarles2018} & $c_1 + c_2\mu^{1/3} + c_3\mu^{2/3}$             & $3$ & $51$ & $3\times10^{-2}$ \\
This work & $c_1 + \frac{1}{\mu + c_2} + \mu^{c_3} - c_4^3$ & $4$ & $50$ & ${O}(10^{-3})\textsuperscript{A}$ \\
\enddata
\tablecomments{A: See Table \ref{Tab: acr_coeff} for the coefficients and errors of our best-fit models.}
\end{deluxetable*}

\begin{deluxetable*}{lrrrrlc}[t]
\tablecaption{Models of the location of stability changing boundaries for planar, circumbinary trajectories based on dynamical systems results in the CR3BP. Using these coefficients in Equation \ref{Eq: acr_fit} produces the fitted curves to the results shown in Figures \ref{fig:acr_pro} and \ref{fig:acr_ret}. \label{Tab: acr_coeff}}
\tablehead{\colhead{Line} & \colhead{$c_1$} & \colhead{$c_2$} & \colhead{$c_3$} &\colhead{$c_4$} & \colhead{Domain} & \colhead{Fractional Error, $\bar{\sigma}$}} 
\startdata
\hline
Prograde & & & & & & \\
\hline
Innermost Stable Trajectory  & $0.53607$ & $ 1.03820$ & $ 0.47113$ & $-0.45708$ & $0.01 \leq \mu \leq  0.50$ & $3.1 \times 10^{-3} $ \\
Exclusion Zone Inner Critical Trajectory & $ 1.23903$ & $ 1.19962$ & $ 1.32271$ & $-0.96885$ & $0.01 \leq \mu \leq  0.50$ & $9.2 \times 10^{-4} $\\
Exclusion Zone Outer Critical Trajectory & $ 0.79351$ & $ 0.79290$ & $ 0.68747$ & $-0.66265$ & $0.01 \leq \mu \leq  0.50$ & $8.9 \times 10^{-4} $\\
\hline
\hline
Retrograde & & & & & & \\
\hline
Innermost Stable Trajectory\textsuperscript{A} & $-0.98016$ & $ 0.83811$ & $ 0.29942$ & $-0.45159$ & $ 0.13 \leq \mu \leq 0.32$ & $3.4 \times 10^{-4} $ \\
Exclusion Zone Inner Critical Trajectory & $0.35382$ & $ 3.71797$ & $ 4.11392$ & $ 3.90066$ & $ 0.15 \leq \mu \leq 0.50 $ & $6.2 \times 10^{-3} $\\
Exclusion Zone Outer Critical Trajectory & $-0.26963$ & $ 0.77862$ & $ 0.80658$ & $-5.89374$ & $ 0.01 \leq \mu \leq 0.32 $ & $3.5 \times 10^{-3} $\\
\enddata
\tablecomments{A: Innermost stable trajectory only within domain applicable to the model. For $\mu < 0.15$, the Exclusion Zone Outer Critical Trajectory is the system's innermost stable retrograde trajectory.}
\end{deluxetable*}

Quantitatively, the \citet{Quarles2018} and our N-body simulations support what we have computed through dynamical systems theory, i.e., trajectories initialized in bounded regions (interior and exterior to the exclusion zone) survive with higher likelihood than those within unbounded regions (the exclusion zone and interior to the tangent bifurcation). However, with perturbations growing by a factor of $e$ over only $12.5-150 \, P_\textrm{bin}$, the question remains as to why any trajectories would exist in the exclusion zone at all. A likely answer falls within two factors (1) dynamical differences between CR3BP and N-body dynamics (2) semi-major axis is an oversimplified parameter for characterizing the unstable solution space between the period-doubling bifurcations. A partial remedy to these complications is to classify the exclusion zone as a range of orbital energies that possess an unstable solution space. In the CR3BP, the Jacobi Constant is invariant along any trajectory and strictly establishes the solution space accessible to the particle. For instance, when we compute that a periodic trajectory possesses locally unstable dynamics, this characterization applies to trajectories in the neighborhood of the limit cycle and hence the same Jacobi Constant. In models considering planet mass and binary eccentricity, a bounded trajectory's Jacobi Constant oscillates; however, assessing the upper and lower limits of the oscillatory behavior may guarantee the trajectory will remain at Jacobi Constants with stable solution spaces. It is feasible and expected for trajectories with Jacobi Constants associated with oscillatory linear dynamics to have semi-major axes that fall within the exclusion zone. In future analysis, it may be beneficial to parameterize regions of stability by the Jacobi Constant for accuracy and consistency across models, along with intuitive measures such as semi-major axis. 

\subsection{A Parametric Fit to Dynamical Results} \label{Sec: parfit}

The dynamical systems analysis provides precise locations for the innermost stable trajectory, $\acr$, and the exclusion zone based on bifurcation analysis but is computationally intensive. For the purpose of generalizing the locations of $\acr$ and the period-doubling exclusion zone to any value of $\mu$ in a circular binary, we would like to fit a simple parametric model to our dynamical results. Prior studies have opted for power series expansions as a basis for their models, as shown in Table \ref{tab: acr_fits}. These parametric models are included in Figures \ref{fig:acr_pro} and \ref{fig:acr_ret} where available, and the goodness of the fit of the parametric model to the dynamical results varies significantly. For instance, \citet{Quarles2018}'s use of the augmented power series expansion to fit $\acr$ is visually and quantitatively a better fit than \citet{Holman1999}'s model. We choose a functional form that combines multiple basis functions,
\begin{equation} \label{Eq: acr_fit}
    a(\mu) = c_1 + \frac{1}{\mu + c_2} + \mu^{c_3} + c_4^3,
\end{equation}
where $\{c_1,...c_4\}$ are real coefficients.  This relation has only one extra free parameter compared to previous studies, but improves the degree-of-freedom weighted fractional error,  

\begin{equation} \label{Eq: error_func}
\bar{\sigma} = \sqrt{\frac{1}{N-k} \sum_{i = 1}^N \frac{(O_i - E_i)^2}{E_i^2}},
\end{equation}

\begin{figure*}[t]
    \centering
    \includegraphics[width = \linewidth]{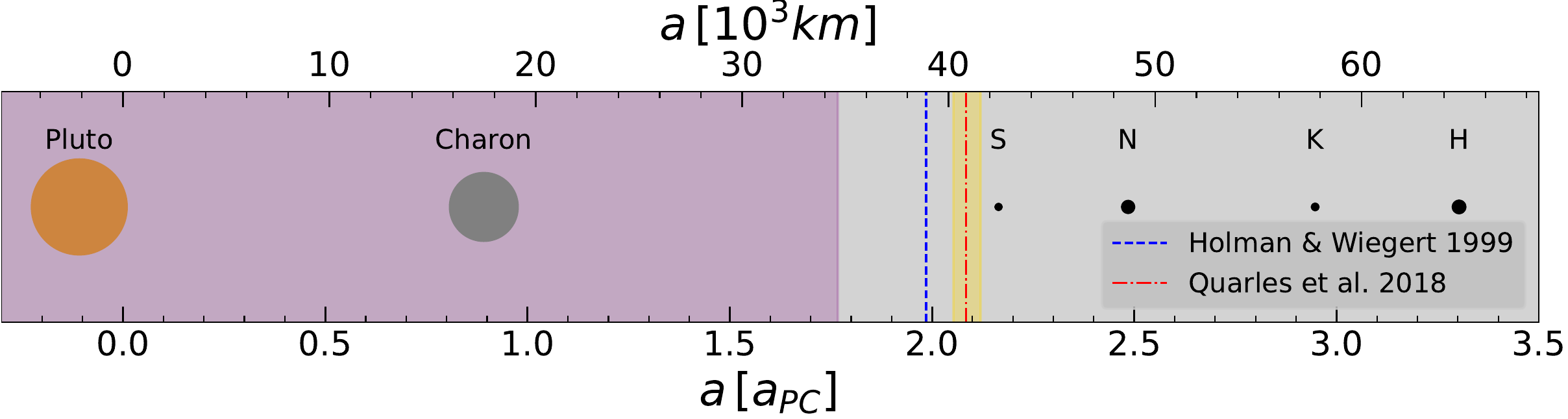}
    \caption{Dynamical architecture of the Pluto-Charon system ($\mu = 0.108$, $e = 0$). Regions of dynamical instability predicted by Floquet theory are colored purple (the region interior to the tangent bifurcation) and yellow (the period-doubling exclusion zone). The innermost stable orbits predicted by parametric models from prior investigations are shown by dashed vertical lines. The innermost satellite, Styx, is located $0.045 \, a_\textrm{PC}$ ($\sim 900 \, \textrm{km}$) away from the outer edge of the exclusion zone, $a_\textrm{Styx} = 1.021 \, a_\textrm{c, EZ}$. No satellites are known in the interior stable region between the exclusion zone and the innermost stable trajectory (purple). }
    \label{fig:Pluto-Charon}
\end{figure*}

by $1-2$ orders of magnitude. In Equation \ref{Eq: error_func}, $N$ is the number of mass ratios, indexed $i$, probed in our dynamical model, $k$ is the number of model coefficients, $E_i$ are the dynamical results at each mass ratio, and $O_i$ are the values computed from our best fit to the model. Using Equation \ref{Eq: acr_fit} with the listed coefficients in Table \ref{Tab: acr_coeff}, one can approximate the geometric semi-major axis of the circumbinary critically stable trajectories in the CR3BP as seen in Figures \ref{fig:acr_pro} and \ref{fig:acr_ret}. To achieve a more precise prediction of the bifurcating trajectories, we suggest either interpolating between the \href{https://doi.org/10.5281/zenodo.7532982}{computed solutions} with a cubic spline or computing desired trajectories directly with the open-source and documented Python package \href{https://github.com/alangfor/pyraa}{\texttt{pyraa}}.

\subsection{Preliminary Application: Pluto-Charon} \label{Sec: PC}

While the motivating celestial objects for this investigation are circumbinary exoplanets, the significant dynamical role of binary eccentricity limits the applicability of our current computations. However, the tidally-locked Pluto-Charon system ($\mu = 0.10854$, $e = 5\times10^{-6}$ \citealt{Brozovic2015}) offers an opportunity to leverage our findings from dynamical systems theory to a well-studied circumbinary system within the Solar System \citep{Weaver2006, Showalter2011, Showalter2012}. The Pluto-Charon system likely formed through an oblique giant impact between proto-planetary debris in the Kuiper belt \citep{Canup2005, Canup2011}. The initial binary orbit properties of Pluto-Charon are unknown; however, tidal forces both increased the semi-major axis, $a_\textrm{PC}$, by $\sim4$ times and circularized the orbit from $e\sim 0.5$ within $1-10$ Myr during the period of tidal expansion \citep{Canup2011, WalshLevison2015}. Theories on the formation of Pluto-Charon's co-planar, dynamically cool satellites, Styx, Nix, Kerberos, and Hydra are divided between pre- and post-tidal evolution \citep{KenyonBromley2021}. Pre-tidal evolution theories suggest the formation of circumbinary satellites among the debris of the giant impact. In this scenario, the satellites would have migrated to the current locations over time through resonance trapping and collisional dampening \citep{WardCanup2006, Cheng2014, WooLee2018, KenyonBromley2021}. Because of difficulties in reproducing the outward migration of the satellites, post-tidal evolution theories invoke a trans-Neptunian object (TNO) colliding with Charon to form a second debris field that formed satellites in their current locations \citep{BromleyKenyon2020}. In either scenario, the innermost orbital debris would have been shaped by Pluto-Charon's critically stable circumbinary regions. Note that our existing CR3BP calculations do not include tidal dissipation.

Pluto-Charon's innermost satellite, Styx, lies on a near-circular, $a_\textrm{Styx} = 2.164 \, a_\textrm{PC}$ trajectory \citep{Brozovic2015}. Styx is near the 3:1 orbital resonance with Pluto-Charon and close to \citet{Holman1999}'s critically stable semi-major axis, $\acr$ \citep{BromleyKenyon2020, BromleyKenyon2021}. Notably, the mass ratio of Pluto-Charon falls within a parameter regime where \citet{Holman1999} and \citet{Quarles2018} polynomial models for $\acr$ fall within the exclusion zone and poorly fit the empirical simulation results (Figure \ref{fig:acr_pro}). From the N-body experimentation results shown in Figure \ref{fig: acr_hist}, we know that dynamically cool circumbinary trajectories are unlikely to survive within the semi-major axis bounds of the exclusion zone. This inaccuracy in prior modeling highlights one advantage of precisely resolving dynamical structures across parameter space through the dynamical systems approach. If the outer edge of the exclusion zone $(a_\textrm{c, EZ} = 2.119 \, a_\textrm{PC})$ is Styx's true critically stable trajectory, we can compute $a_\textrm{Styx} = 1.021 \, a_\textrm{c, EZ}$ -- $\sim 900 $ km away from the exclusion zone. Based on our computations, a critically stable trajectory is closer to Styx than predicted by the parametric models from \cite{Holman1999} or \cite{Quarles2018}. Styx's current trajectory is unlikely to be within the exclusion because the typical ejection time for trajectories initialized inside the exclusion zone is $<10^3 \, P_\textrm{bin}$, or 20 years.\footnote{Using results from $\mu = 0.10$ N-body simulations shown in Figure \ref{fig: acr_hist}.} In N-body simulations, \citet{KenyonBromley2022} found scenarios in which Styx is ejected after a destabilizing interaction with Nix and the 3:1 resonance, which is consistent with our finding that Styx is near the critical trajectory for stability.

If Styx did form in-situ, its proximity to the exclusion zone likely limited its growth and would be of great interest to theories for circumbinary satellite formation. No satellites have been detected in the interior stable region, between the exclusion zone and the innermost stable trajectory. \citet{KenyonBromley2019a} found that massless particles could survive in this region $(a = 1.7–2.1 \, a_\textrm{PC})$ for 10 Myr. This may indicate clearing processes in the Pluto-Charon system's evolutionary history or evidence that the interior stable region is not a robust environment for debris agglomeration. Imaging with JWST opens a new parameter space for new possible satellite and particle debris detection \citep{KenyonBromley2019a}.

\section{Implications of CR3BP Emergent Dynamical Structures for Planet Formation} \label{Sec: DST4PF}

In this investigation, we have identified and computed two dynamical structures, an innermost stable trajectory and a dynamical exclusion zone, that are emergent from the CR3BP and may be relevant to the architectures of observed planets in close binary systems. Since the location of the dynamical exclusion zone is in better agreement with prior investigations of the innermost stable trajectory in the CR3BP, one might wonder if it persists in real physical systems -- where stars orbits' have non-zero eccentricities and planets have finite masses. We suspect that the period-doubling exclusion zone persists when these criteria are relaxed for several reasons: (1) In dynamical systems, emergent structures (such as the exclusion zone) tend to vary smoothly as model parameters (such as eccentricity) change \citep{Boudad2022}. (2) The \citet{Quarles2018} simulations of massive planets around \textit{eccentric} binary systems consistently produced a gap in surviving trajectories similar to those present in Figure \ref{fig: acr_hist}, suggesting that the period-doubling exclusion zone is not unique to the case of circular binaries or test particle trajectories. (3) The occurrence of Kepler CBPs near $\acHW$ is statistically significant even when observational biases are considered \citep{WinnFabrycky2015, Li2016}.

The location of critically stable trajectories likely influences how and where circumbinary planets may form and be observed around close binaries \citep{Thun2018}. Hydrodynamical simulations of gaseous and grain-filled protoplanetary disks produce central cavities in the disk near $\acHW$ where trajectories around the binary are unstable \citep{Coleman2022}. The central cavity results in a pressure gradient within the disk and the accumulation of rocky material near its inner edge. The Kepler CBPs observed near $\acHW$ will have formed either (1) in-situ near the cavity edge or (2) migrated through the disk and parked near the cavity edge. Analysis of the planet-forming environment disfavors in-situ formation, but hydrodynamical simulations have difficulty parking simulated planets near their observed locations \citep{Paardekooper2012, Thun2018, Coleman2022}. 

The plausible existence of two separate regions of stability around a binary both complicates and places constraints on theories of CBP planet formation. The non-detection of CBPs interior to $\acHW$, and thus the exclusion zone, suggests planets cannot traverse the unstable region before being ejected or form in-situ within the stable region between the exclusion zone and the innermost stable trajectory. In this case, the outermost critically stable trajectory of the exclusion zone would serve as $\acr$ for planets migrating from beyond the snow lines. Quantifying the rarity of planets between the innermost stable trajectory and the period-doubling exclusion zone will require (1) careful completeness correction via injection/recovery in the manner of \citet{Armstrong2014} and (2) an accurate determination of the extent of the period-doubling exclusion zone for eccentric binary stars. These next steps are challenging but will likely clarify the dynamical conditions in which CBPs form and evolve. 

\section{Remarks on Future Applications}

The techniques and results presented in this manuscript are a first attempt at applying modern dynamical systems theory to pertinent questions in dynamical astronomy. We have successfully enhanced the findings of prior theoretical investigations of circumbinary dynamics around a circular binary. However, a few challenges still remain to be solved for broad application to physical systems -- namely, extension to eccentric binaries and connection to observable quantities. Below, we remark on how these next steps may proceed.

\subsection{Extension to Eccentric Binaries}

The host stars of most circumbinary planets are on sufficiently eccentric orbits where the binary star eccentricity dominates the stability of the system \citep{Kostov2021}. To directly compare the predictions of dynamical systems theory to observed circumbinary planets, it is necessary to generalize the calculations for motion around an eccentric binary. The Elliptical Restricted Three-body Problem (ER3BP) provides the appropriate extension to the CR3BP for modeling a massless particle in the gravitational environment of an eccentric binary. The ER3BP is a more challenging dynamical system than the CR3BP as a consequence of the non-autonomous (time-varying) equations of motion in the synodic frame \citep{Hiday1992}. Recall \S \ref{Sec: PerSol} that continuous families of periodic solutions are present in the CR3BP due to the autonomous (time-independent) equations of motion. While isolated periodic solutions exist in the ER3BP at resonances with the synodic equations of motion, continuous families are required to resolve dynamical structures and stability bifurcations. Therefore, an extension of the dynamical systems theory approach to eccentric binaries requires computing continuous families of 2-D quasi-periodic orbits (QPOs) that take the place of periodic trajectories in the non-autonomous model. Numerical methods for computing QPOs in restricted dynamical models and their applications to astrodynamics is an active area of study \citep{OlikaraScheeres2012, BaresiOlikaraScheeres2018, McCarthyHowell2021, McCarthyHowell2022}. The computation and analysis of 2-D QPOs in the ER3BP will yield the appropriate generalization of the dynamical systems approach to CBPs in eccentric binaries.

\subsection{Connection to Observable Quantities}

In order to leverage the precise stability predictions from Floquet theory, the outputs of these calculations, limit cycles, must be connected with observable quantities. Our existing analysis provides parametric models that capture the geometric semi-major axis of stability-changing periodic trajectories as a function of a circular binary's mass ratio (Tables \ref{tab: acr_fits} and \ref{Tab: acr_coeff}). Using an observed system's mass ratio, one could calculate its critically stable semi-major axes and compare them against known circumbinary orbits in the system (see \S \ref{Sec: PC}). Recognizing that periodic trajectories are idealized solutions, future work must verify the appropriate set of parameters, e.g., $a_\mathrm{geo},\ C_J,\ e_\mathrm{geo}$ that accurately generalizes predictions for the stability of non-periodic trajectories in higher-fidelity gravitational models. The future applications of Floquet stability may include objects other than CBPs such as debris disks or Solar System objects. In each application, the measurable quantities should have a reasonably straightforward connection with stability computations from Floquet theory. For CBPs, the orbital solutions are computed as parameters satisfying constraints in a photo-dynamical model (see \citealt{Kostov2016} and references within). From these orbital solutions, instantaneous Keplerian orbital elements and a Jacobi Constant can be calculated. Similarly, Solar System objects typically have well-constrained orbital solutions. \cite{LeePeale2006}'s first-order epicyclic theory may also serve as a useful bridge between measurable quantities and dynamical system's outputs by extracting orbital parameters through \citet{WooLee2020}'s Fast Fourier Transform method. Measurable orbital parameters from a circumbinary debris disk are more likely to be limited to a distance and uncertainty from the stellar source(s) based on an SED \citep{Farihi2017}.

\section{Summary}

Motivated by the architectures of observed circumbinary planetary systems, we examined the applicability of a modern dynamical systems theory approach to better understand the orbital dynamics of CBPs. Our approach leveraged the computation of periodic families of planar circumbinary trajectories for mass ratios $\mu \in [0.01, 0.50]$. The sets of circumbinary retrograde and prograde periodic trajectories are useful for (1) characterizing the solution space of guaranteed bounded trajectories in the three-body environment (2) analyzing the local stability properties and emergent dynamical structures of circumbinary orbits. 

Bifurcations in the linear stability of periodic families determined the innermost stable trajectories for a given mass ratio system. For retrograde orbits, this trajectory exists as the maximum-valued Jacobi Constant (minimum energy) limit cycle within the family and ranges from $a_\textrm{geo} \in [1.02, 0.52]$ for binary mass ratios $\mu \in [0.01, 0.50]$. The prograde innermost stable trajectories range from $a_\textrm{geo} \in  [1.61, 1.85]$ for $\mu \in [0.01,0.50]$. Our bifurcation analysis of the periodic families has provided exact and reproducible calculations of $\acr$ that are interior to previously determined values.

Our bifurcation analysis also yielded a region of unstable trajectories, which we call the period-doubling exclusion zone, that is exterior to the innermost stable trajectory. The exclusion zone's solution space, bounded between a minimum and maximum Jacobi constant, is disconnected from the unstable solution space interior to the innermost stable trajectory. Analysis of the prograde periodic trajectories' sidereal periods found the $3:1$ mean-motion resonance resides within the exclusion zone at all mass ratios, while no other critically stable trajectory has a distinct connection to a resonance. Previous determinations for the location of the innermost stable trajectories reside outside the critically stable trajectories of the exclusion zone. At the mass ratios of Kepler CBPs, prior locations of the innermost stable trajectories are slightly exterior to the exclusion zone and their associated models inform the statistical pile-up of CBPs. The $\sim 0.1 \, a_\textrm{bin}$ wide region of the exclusion zone is positioned near $2.1 \, a_\textrm{bin}$, a location that corresponds to a deficit in surviving trajectories for Jupiter-mass planet orbiting a circular binary for $10^5 \, P_\textrm{bin}$. Our interpretation of this result includes the possibility that CBPs are found near the outer boundary of a dynamical exclusion zone, rather than the system's innermost stable trajectory. Further application of the dynamical systems approach to eccentric binaries and long-term N-body integrations will assist in determining if the exclusion zone is persistent in physical systems.  

The dynamical systems theory approach has uncovered regions of stable and unstable trajectories in the CR3BP that were not readily identifiable through numerical simulation or analytic perturbation methods. Applying modern dynamical systems techniques to other applicable cases in dynamical astronomy may yield novel insights into the emergent dynamical structures influencing astrophysical processes in multi-body gravitational environments. 

\textbf{Acknowledgements}
We thank F. Adams and J. Becker for their helpful suggestions throughout this investigation. We also thank M. Wyatt for discussions that improved the quality of this work. A.L. thanks the members of the Purdue Multi-Body Dynamics research group for offering technical guidance on applications of dynamical systems theory to the CR3BP. We appreciate the quality feedback and recommendations from the anonymous referee, including the suggestion of a preliminary application to the Pluto-Charon system. This research was supported in part by the Notre Dame Center for Research Computing. This material is based upon work supported by the National Science Foundation Graduate Research Fellowship Program under Grant No. DGE-1842166.

\software{\texttt{pyraa}, \texttt{rebound} \citep{Rein2012}, \texttt{scipy} \citep{Virtanen2020}, \texttt{numpy} \citep{Numpy2020}, \texttt{numba} \citep{numba2015}, \texttt{matplotlib} \citep{matplotlib2007}, \texttt{TikZiT}}

\bibliography{references.bib}

\end{document}